\documentclass[aps,showpacs]{revtex4}
\begin{document}
\draft
\title{Spurious modes in extended RPA theories}
\author{Mitsuru Tohyama}
\address{Kyorin University School of Medicine, 
Mitaka, Tokyo 181-8611, Japan}
\author{Peter Schuck}
\address{
Institut de Physique Nucl$\acute{e}$aire, IN2P3-CNRS, Universit$\acute{e}$ Paris-Sud, F-91406 Orsay Cedex, France}
\date{\today}
\begin{abstract}
Necessary conditions that the spurious state associated with
the translational motion and its double-phonon state have zero
excitation energy in extended RPA (ERPA) theories which include both one-body and two-body amplitudes 
are investigated
using the small amplitude limit of 
the time-dependent density-matrix theory (STDDM).
STDDM provides us with a quite general form of ERPA as compared with other
similar theories in the sense that 
all components of one-body and two-body amplitudes
are taken into account.
Two conditions are found necessary to guarantee the above property
of the single and double spurious states: 
The first is that no truncation in the
single-particle space should be made. This
condition is necessary for the closure relation to be used and is common for the single and double
spurious states.
The second depends on the mode. For the single
spurious state all components of the one-body amplitudes must be included, and for the 
double spurious state all components of one-body and two-body amplitudes have to be included. 
It is also shown that the Kohn theorem and the continuity equations for transition densities and
currents hold under the same conditions as the spurious states.
ERPA theories 
formulated using the Hartree-Fock ground state have a non-hermiticity problem.
A method for formulating ERPA with hermiticity is also proposed
using the time-dependent density-matrix formalism.
\end{abstract}
\maketitle

\section{Introduction}

The double phonon states of giant resonances have become
the subject of a number of recent experimental and theoretical
investigations\cite{Chomaz,Bertulani}.
In the case of giant resonances (single phonon states), the random phase approximation (RPA)
has extensively been used as a standard microscopic theory to
study basic properties of giant resonances \cite{Tsai}.
It is guaranteed in RPA that physical states do not couple to spurious
states such as that associated with the translational motion 
because RPA in the Hartree-Fock (HF) basis gives zero excitation energy to spurious states.
For a microscopic study of the double phonon states of giant resonances, we need to extend
RPA to deal with two-body amplitudes as well as one-body amplitudes.
One of such an extended RPA theory (ERPA) may be the Second RPA (SRPA) \cite{Saw} which
has also extensively been used to study decay properties of giant resonances \cite{Yann,Wam}.
When the double phonon states are studied in ERPA, 
it should be guaranteed that both spurious states and their double-phonon
states are decoupled from physical states.
The aim of this paper is to investigate 
necessary conditions that the spurious state associated with
the translational motion and its double-phonon state have zero
excitation energy in ERPA.
We use the small amplitude limit of the time-dependent density-matrix theory (STDDM) \cite{TG1}.
The reason why STDDM is used
is that containing all components of one-body and two-body amplitudes,
STDDM constitutes a more general framework of ERPA than SRPA.
It will be shown that keeping all components of the one-body and two-body
amplitudes in ERPA is essential in bringing the spurious states to zero excitation energy.
Any ERPA including STDDM , which is formulated using an approximate
ground state, inherently has asymmetry and non-hermiticity.
A method for recovering symmetry and hermiticity in the framework of
the time-dependent density-matrix formalism is also proposed
in this paper.
The paper is organized as follows: STDDM is presented
and its relation to other ERPA formulations is discussed in Sect.2.
Necessary conditions to give zero excitation energy to the spurious
state associated with the translational motion and its double-phonon state
are discussed in Sect.3. The Kohn theorem \cite{Kohn,Brey,Dobson} and the continuity
equations for transition densities and currents are also discussed as related subjects in Sect.3.
In Sect.4 a method for formulating ERPA with hermiticity is proposed and Sect.5 is devoted to a summary.

\section{Extended RPA formalism}
\subsection{Small amplitude limit of the time-dependent density-matrix theory}

The time-dependent density-matrix theory (TDDM)
gives the time-evolution of
a one-body density-matrix $\rho(1,1')$ and the correlated part $C(12,1'2')$ 
of a two-body 
density-matrix \cite{WC}, where numbers denote space, spin, and isospin coordinates.
Linearizing the equations of motion for $\rho$ and $C$ with respect to
$\delta\rho$ and $\delta C$, where $\delta\rho$ and $\delta C$
denote deviations from the ground-state values $\rho_0$ and $C_0$ i.e.
$\delta\rho=\rho-\rho_0$ and
$\delta C=C-C_0$, respectively, we obtain STDDM\cite{TG1}.
Expanding
$\delta\rho$ and $\delta C$ with single-particle states
$\psi_{\alpha}$ as
\begin{eqnarray}
\delta\rho(11',t)&=&\sum_{\alpha\alpha'}x_{\alpha\alpha'}(t)\psi_{\alpha}(1,t)
\psi_{\alpha'}^{*}(1',t), 
\end{eqnarray}
\begin{eqnarray}
\delta C(121'2',t)
=\sum_{\alpha\beta\alpha'\beta'}X_{\alpha\beta\alpha'\beta'}(t)
\psi_{\alpha}(1,t)\psi_{\beta}(2,t)
\psi_{\alpha'}^{*}(1',t)\psi_{\beta'}^{*}(2',t),
\end{eqnarray}
and assuming the HF ground state, that is, $\rho_0$
is the one-body density-matrix
in HF approximation and $C_0=0$,
we obtain the following equations of STDDM 
for the Fourier components of $x_{\alpha\alpha'}(t)$
and $X_{\alpha\beta\alpha'\beta'}(t)$ \cite{TG1}:
\begin{eqnarray}
(\omega - \epsilon_{\alpha}+\epsilon_{\alpha'})x_{\alpha\alpha'}
&=& (f_{\alpha'}-f_{\alpha})\sum_{\lambda\lambda'}\langle\alpha\lambda|v|\alpha'\lambda'\rangle_A
x_{\lambda'\lambda} \nonumber \\
&+& \sum_{\lambda\lambda'\lambda''}[X_{\lambda\lambda'\alpha'\lambda''}
\langle\alpha\lambda''|v|\lambda\lambda'\rangle
- X_{\alpha\lambda'\lambda\lambda''}
\langle\lambda\lambda''|v|\alpha'\lambda'\rangle],
\label{stddm1}
\end{eqnarray}
\begin{eqnarray}
(\omega &-& \epsilon_{\alpha} - \epsilon_{\beta}+\epsilon_{\alpha'}
+\epsilon_{\beta'})X_{\alpha\beta\alpha'\beta'} =
-\sum_{\lambda}[(\bar{f_{\beta}}f_{\alpha'}f_{\beta'}
+f_{\beta}\bar{f}_{\alpha'}\bar{f}_{\beta'})
\langle\lambda\beta|v|\alpha'\beta'\rangle_Ax_{\alpha\lambda}
\nonumber \\
&+&(\bar{f_{\alpha}}f_{\alpha'}f_{\beta'}+f_{\alpha}\bar{f}_{\alpha'}
\bar{f}_{\beta'})\langle\alpha\lambda|v|\alpha'\beta'\rangle_Ax_{\beta\lambda}
-(\bar{f_{\alpha}}\bar{f_{\beta}}f_{\beta'}
+f_{\alpha}f_{\beta}\bar{f}_{\beta'})
\langle\alpha\beta|v|\lambda\beta'\rangle _A x_{\lambda\alpha'}
\nonumber \\
&-&(\bar{f_{\alpha}}\bar{f_{\beta}}f_{\alpha'}+f_{\alpha}f_{\beta}
\bar{f}_{\alpha'})\langle\alpha\beta|v|\alpha'\lambda\rangle_Ax_{\lambda\beta'}]
\nonumber \\
&+&\sum_{\lambda\lambda'}[(1-f_{\alpha}-f_{\beta})
\langle\alpha\beta|v|\lambda\lambda'\rangle X_{\lambda\lambda'\alpha'\beta'}
-(1-f_{\alpha'}-f_{\beta'})\langle\lambda\lambda'|v|\alpha'\beta'\rangle
X_{\alpha\beta\lambda\lambda'}] \nonumber \\
&+&\sum_{\lambda\lambda'}
[(f_{\alpha'}-f_{\alpha})
\langle\alpha\lambda|v|\alpha'\lambda'\rangle_AX_{\lambda'\beta\lambda\beta'}
-(f_{\beta'}-f_{\alpha})
\langle\alpha\lambda|v|\lambda'\beta'\rangle_AX_{\lambda'\beta\alpha'\lambda}
\nonumber \\
&+&(f_{\beta'}-f_{\beta})
\langle\lambda\beta|v|\lambda'\beta'\rangle_AX_{\alpha\lambda'\alpha'\lambda}
-(f_{\alpha'}-f_{\beta})
\langle\lambda\beta|v|\alpha'\lambda'\rangle_AX_{\alpha\lambda'\lambda\beta'}],
\label{stddm2}
\end{eqnarray}
where $\epsilon_{\alpha}$ is the HF single-particle energy, $f_{\alpha}=1 (0)$ for occupied (unoccupied) 
single-particle states and $\bar{f_{\alpha}}=1-f_{\alpha}$, and
the subscript $A$ indicates that the corresponding matrix element is antisymmetrized.
Let us mention that
eqs.(\ref{stddm1}) and (\ref{stddm2}) may also be obtained from the following equations of motion
\begin{eqnarray}
\langle\Phi_0|[a^+_{\alpha'}a_{\alpha},H]|\Phi\rangle
&=&\omega\langle\Phi_0|a^+_{\alpha'}a_{\alpha}|\Phi\rangle ,
\label{var1}
\\
\langle\Phi_0|[a^+_{\alpha'}a^+_{\beta'}a_{\beta}a_{\alpha},H]|\Phi\rangle
&=&\omega\langle\Phi_0|a^+_{\alpha'}a^+_{\beta'}a_{\beta}a_{\alpha}|\Phi\rangle,
\label{var2}
\end{eqnarray}
where 
$[~]$ is the commutation relation, $H$ the total hamiltonian 
consisting of the kinetic energy term and a two-body interaction,
$|\Phi_0\rangle$ 
the ground-state wavefunction and
$|\Phi\rangle$ the wavefunction for an excited state with excitation energy $\omega$.
Linearizing eqs.(\ref{var1}) and (\ref{var2})
with respect to $x_{\alpha\alpha'}=\langle\Phi_0|a^+_{\alpha'}a_{\alpha}|\Phi\rangle$
and $X_{\alpha\beta\alpha'\beta'}=
\langle\Phi_0|a^+_{\alpha'}a^+_{\beta'}a_{\beta}a_{\alpha}|\Phi\rangle$, and
assuming the HF ground state for $|\Phi_0\rangle$ when expectation values for
the ground state are evaluated such as 
$\langle\Phi_0|a^+_{\alpha}a_{\alpha'}|\Phi_0\rangle
\approx\delta_{\alpha\alpha'}f_{\alpha}$, we obtain eqs.(\ref{stddm1}) and (\ref{stddm2}).

In the following we discuss some relations of STDDM with RPA and other versions of 
ERPA. When the coupling to
the two-body amplitudes $X_{\alpha\beta\alpha'\beta'}$ is neglected in eq.(\ref{stddm1}), the equation
for the one-body amplitudes becomes
\begin{eqnarray}
(\omega - \epsilon_{\alpha}+\epsilon_{\alpha'})x_{\alpha\alpha'}
&=& (f_{\alpha'}-f_{\alpha})\sum_{\lambda\lambda'}\langle\alpha\lambda|v|\alpha'\lambda'\rangle_A
x_{\lambda'\lambda}. 
\label{grpa}
\end{eqnarray}
When $f_\alpha$ is the Fermi-Dirac distribution, 
eq.(\ref{grpa}) is equivalent to the finite temperature RPA \cite{Vaut,Somm}.
Hereafter single-particle indices $p$ and $h$ are used to refer to unoccupied and occupied single-particle 
states, respectively. 
Since the sums on the right-hand sides of the equations for $x_{ph}$ and $x_{hp}$ are
unrestricted, $x_{ph}$ and $x_{hp}$ can couple to $x_{pp'}$ and $x_{hh'}$. 
Such a coupling scheme of eq.(\ref{grpa}) may be better understood in matrix form
\begin{eqnarray}
\left(
\begin{array}{cccc}
 \epsilon_{p}-\epsilon_{h}+\langle p h'|v|h p'\rangle_A & \langle p p'|v|h h'\rangle_A &
\langle p p'|v|h p''\rangle_A & \langle p h'|v|h h''\rangle_A\\
-\langle h h'|v|p p'\rangle_A & \epsilon_{h}-\epsilon_{p}-\langle h p'|v|p h'\rangle_A &
-\langle h p'|v|p p''\rangle_A & -\langle h h'|v|p h''\rangle_A \\
0 & 0& \epsilon_{p}-\epsilon_{p'} & 0 \\
0 & 0& 0& \epsilon_{h}-\epsilon_{h'}\\
\end{array}
\right)
\left(
\begin{array}{c}
x_{p'h'} \\
x_{h'p'} \\
x_{p''p'}\\
x_{h''h'}
\end{array}
\right)
=\omega
\left(
\begin{array}{c}
x_{ph} \\
x_{hp} \\
x_{pp'}\\
x_{hh'}
\end{array}
\right) ,
\end{eqnarray}
where obvious summation symbols and Kronecker's $\delta$'s are omitted for simplicity.
Since the hamiltonian matrix is non-hermitian, $x_{\alpha\alpha'}$ is orthogonal not to $x_{\alpha\alpha'}$
but to a left-hand-side eigenvector $\tilde{x}_{\alpha\alpha'}$ which satisfies
\begin{eqnarray}
(\omega - \epsilon_{\alpha}+\epsilon_{\alpha'})\tilde{x}^*_{\alpha\alpha'}
&=& \sum_{\lambda\lambda'}(f_{\lambda'}-f_{\lambda})\langle\lambda\alpha'|v|\lambda'\alpha\rangle_A\tilde{x}^*_{\lambda\lambda'}
\nonumber \\
&=&\sum_{ph}(\langle p\alpha'|v|h \alpha\rangle_A
\tilde{x}^*_{ph}-\langle h\alpha'|v|p \alpha\rangle_A
\tilde{x}^*_{hp})~~({\rm at ~temperature}~T=0).
\label{lgrpa}
\end{eqnarray}
The matrix form of eq.(\ref{lgrpa}) becomes
\begin{eqnarray}
&(&
%\begin{array}{c}
\tilde{x}^*_{p''h''},
\tilde{x}^*_{h''p''}, 
\tilde{x}^*_{pp'},
\tilde{x}^*_{hh'}
%\end{array}
)
\left(
\begin{array}{cccc}
 \epsilon_{p}-\epsilon_{h}+\langle p'' h|v|h'' p\rangle_A & \langle p'' p|v|h'' h\rangle_A &
\langle p'' p|v|h'' p'\rangle_A & \langle p'' h|v|h'' h'\rangle_A\\
-\langle h'' h|v|p'' p\rangle_A & \epsilon_{h}-\epsilon_{p}-\langle h'' p|v|p'' h\rangle_A &
-\langle h'' p|v|p'' p'\rangle_A & -\langle h'' h|v|p'' h'\rangle_A \\
0 & 0& \epsilon_{p}-\epsilon_{p'} & 0 \\
0 & 0& 0& \epsilon_{h}-\epsilon_{h'}
\end{array}
\right)
\nonumber \\
&=&\omega
\left(
%\begin{array}{c}
\tilde{x}^*_{ph}, 
\tilde{x}^*_{hp}, 
\tilde{x}^*_{pp'},
\tilde{x}^*_{hh'}
%\end{array}
\right).
\end{eqnarray}
The ortho-normal condition is written as
\begin{eqnarray}
\langle\tilde{\lambda}|\lambda'\rangle&=&
\sum_{\alpha\alpha'}\tilde{x}^*_{\alpha\alpha'}(\lambda)x_{\alpha\alpha'}(\lambda')
=\delta_{\lambda\lambda'},
\end{eqnarray}
where $|\lambda\rangle$ represents an eigenvector $x_{\alpha\alpha'}$
with the eigenvalue $\omega_{\lambda}$, and $|\tilde{\lambda}\rangle$ the left-hand eigenvector of
the hamiltonian matrix with the eigenvalue $\omega_{\lambda}$.
The completeness relation becomes
\begin{eqnarray}
\sum_{\lambda}|\lambda\rangle\langle\tilde{\lambda}|=\sum_{\lambda}
x_{\alpha\alpha'}(\lambda)
\tilde{x}^*_{\beta\beta'}(\lambda)
=I,
\end{eqnarray}
where $I$ is the unit matrix. These ortho-normal and completeness relations are generalizations of the RPA ones.
Due to the occupation factor $f_\alpha-f_{\alpha'}$
the one-body amplitudes $x_{pp'}$ and $x_{hh'}$ vanish unless $\omega =\epsilon_{\alpha}-\epsilon_{\alpha'}$ (see eq.(\ref{grpa}))
whereas $\tilde{x}_{\alpha\alpha'}$ always have all components as seen from eq.(\ref{lgrpa}):
$\tilde{x}_{\alpha\alpha'}$ corresponds to the generalized RPA
amplitude which appears in the Landau's expression for the damping width of zero sound \cite{Ando,Adachi}.
If the particle (p) - particle (p) and hole (h) - hole (h) components
of $x_{\alpha\alpha'}$ are neglected, eq.(\ref{grpa}) is reduced to the RPA equations,
\begin{eqnarray}
(\omega - \epsilon_{p}+\epsilon_{h})x_{ph}
&=& \sum_{p'h'}[
  \langle ph'|v|hp'\rangle_Ax_{p'h'} 
 + \langle pp'|v|hh'\rangle_Ax_{h'p'}], 
\label{rpa1}\\
(\omega - \epsilon_{h}+\epsilon_{p})x_{hp} 
&=-& \sum_{p'h'}[
 \langle hh'|v|pp'\rangle_Ax_{p'h'}
 + \langle hp'|v|ph'\rangle_Ax_{h'p'}].
\label{rpa2}
\end{eqnarray}
When the coupling to the one-body amplitudes is neglected in eq.(\ref{stddm2}), the equation for
the two-body amplitudes become 
\begin{eqnarray}
(\omega &-& \epsilon_{\alpha} - \epsilon_{\beta}+\epsilon_{\alpha'}
+\epsilon_{\beta'})X_{\alpha\beta\alpha'\beta'} =
\sum_{\lambda\lambda'}[(1-f_{\alpha}-f_{\beta})
\langle\alpha\beta|v|\lambda\lambda'\rangle X_{\lambda\lambda'\alpha'\beta'}
-(1-f_{\alpha'}-f_{\beta'})\langle\lambda\lambda'|v|\alpha'\beta'\rangle
X_{\alpha\beta\lambda\lambda'}] \nonumber \\
&+&\sum_{\lambda\lambda'}
[(f_{\alpha'}-f_{\alpha})
\langle\alpha\lambda|v|\alpha'\lambda'\rangle_AX_{\lambda'\beta\lambda\beta'}
-(f_{\beta'}-f_{\alpha})
\langle\alpha\lambda|v|\lambda'\beta'\rangle_AX_{\lambda'\beta\alpha'\lambda}
\nonumber \\
&+&(f_{\beta'}-f_{\beta})
\langle\lambda\beta|v|\lambda'\beta'\rangle_AX_{\alpha\lambda'\alpha'\lambda}
-(f_{\alpha'}-f_{\beta})
\langle\lambda\beta|v|\alpha'\lambda'\rangle_AX_{\alpha\lambda'\lambda\beta'}].
\label{Pawel}
\end{eqnarray}
This equation is equivalent to the formula given in Ref.\cite{Dani} for the two-body space.
Keeping only the 2p-2h, 2h-2p and 1p1h-1p1h components of $X_{\alpha\beta\alpha'\beta'}$
in eq.(\ref{Pawel}) 
leads to the version of ERPA 
for low-lying two-phonon states \cite{Sakata}.
It has been pointed out \cite{Sakata} that the 1p1h-1p1h components of 
$X_{\alpha\beta\alpha'\beta'}$ are important to reproduce collectivity
of low-lying double-phonon states.
A time-dependent version of eq.(\ref{Pawel}) has been applied to the double-phonon states
of giant dipole and quadrupole resonances in $^{40}$Ca
using a realistic Skyrme-type interaction for both the mean-field potential
and the residual interaction, and it was found that
the 2p-2h, 2h-2p and 1p1h-1p1h components are the most important two-body 
amplitudes for these double-phonon states \cite{Toh}.
In eqs.(\ref{stddm1}) and (\ref{stddm2}) the one-body amplitude $x_{\alpha\alpha'}$ and the 
two-body amplitude $X_{\alpha\beta\alpha'\beta'}$ have all components:
For example, $x_{\alpha\alpha'}$ has 1p-1h,
1h-1p, 1p-1p and 1h-1h components. On the other hand
only the 1p-1h and 1h-1p components of $x_{\alpha\alpha'}$ and the 2p-2h and
2h-2p components of $X_{\alpha\beta\alpha'\beta'}$ are taken into account in
SRPA \cite{Saw} and the SRPA equations are obtained from
eqs.(\ref{stddm1}) and (\ref{stddm2}) by keeping only these amplitudes.

Equations (\ref{stddm1}) and (\ref{stddm2}) have asymmetric couplings between the $x_{\alpha\alpha'}$ and 
$X_{\alpha\beta\alpha'\beta'}$ amplitudes:
In eq.(\ref{stddm1}) $x_{\alpha\alpha'}$ couples to all components of 
$X_{\alpha\beta\alpha'\beta'}$, while in eq.(\ref{stddm2}) 
only the 2p-2h, 1p-3h and 1h-3p components of
$X_{\alpha\beta\alpha'\beta'}$
(and their complex conjugates) couple to $x_{\alpha\alpha'}$
due to the occupation factors ($\bar{f_{\beta}}f_{\alpha'}f_{\beta'}$
etc.).
Equations (\ref{grpa}) and (\ref{Pawel}) which have no coupling between one-body and two-body amplitudes are also non-hermitian
due to occupation factors such as $f_{\alpha'}-f_{\alpha}$.
The asymmetry and non-hermiticity originate from the structure of the equations for the
reduced density matrices (see eqs.(\ref{var1}) and (\ref{var2})).
For a non-hermitian hamiltonian matrix, the left-hand-side eigenvectors of the
hamiltonian matrix constitute a basis which is orthogonal to $(x_{\alpha\alpha'}, X_{\alpha\beta\alpha'\beta'})$,
and
the ortho-normal condition is written as
\begin{eqnarray}
\langle\tilde{\lambda}|\lambda'\rangle&=&
\sum_{\alpha\alpha'}\tilde{x}^*_{\alpha\alpha'}(\lambda)x_{\alpha\alpha'}(\lambda')
+\sum_{\alpha\beta\alpha'\beta'}\tilde{X}^*_{\alpha\beta\alpha'\beta'}(\lambda)X_{\alpha\beta\alpha'\beta'}(\lambda')
=\delta_{\lambda\lambda'},
\end{eqnarray}
where $|\lambda\rangle$ represents an eigenvector $(x_{\alpha\alpha'}, X_{\alpha\beta\alpha'\beta'})$
with the eigenvalue $\omega_{\lambda}$, and $|\tilde{\lambda}\rangle$ the left-hand-side eigenvector of
the hamiltonian matrix with the same eigenvalue.
The completeness relation is written as
\begin{eqnarray}
\sum_{\lambda}
\left(
\begin{array}{c}
x_{\alpha\alpha'}(\lambda)\\
X_{\alpha\beta\alpha'\beta'}(\lambda)
\end{array}
\right)
(\tilde{x}^*_{\beta\beta'}(\lambda)~\tilde{X}^*_{\beta\gamma\beta'\gamma'}(\lambda))
=I.
\end{eqnarray}
The asymmetry and non-hermiticity in eqs.(\ref{stddm1}) and (\ref{stddm2}) are necessary
to prove the properties of the spurious states
and the Kohn theorems as will be discussed below. However, due to the non-hermiticity of the problem
some of the eigenvalues may become complex. 
Our exploratory numerical calculations 
for the oxygen isotopes $^{22,24}$O using the neutron $2s$ and $1d$ states and a pairing-type residual interaction
which had been used in the calculations of quadrupole states in these nuclei \cite{2+}
show that the non-hermiticity of STDDM is quite moderate: Only a small fraction 
(about 10\%)
of the eigenstates have complex energies, whose imaginary parts are less than  0.1 MeV.
The results of these numerical calculations will be discussed elsewhere. On the other hand 
we will show in Sect. 4 that there is a prescription for constructing ERPA with symmetry and hermiticity
using a correlated ground state in TDDM.

\section{Single and double spurious states}
\subsection{One-body and two-body operators for the translational motion}
We consider the following 
one-body and two-body operators associated with the translational motion:
\begin{eqnarray}
\vec{P}=\sum_{\alpha\beta}\langle\alpha|-i\nabla|\beta\rangle
a^+_{\alpha}a_{\beta}
\end{eqnarray}
and 
\begin{eqnarray}
\vec{P}\cdot\vec{P}=\sum_{\alpha\alpha'}
\langle\alpha|-\nabla^2|\alpha'\rangle a^+_{\alpha}a_{\alpha'}
+\sum_{\alpha\beta\alpha'\beta'}\langle\alpha|-i\nabla|\alpha'\rangle
\cdot\langle\beta|-i\nabla|\beta'\rangle
a^+_{\alpha}a^+_{\beta}a_{\beta'}a_{\alpha'}.
\end{eqnarray}
Since the hamiltonian $H$ has translational invariance, 
these operators commute with $H$, that is, 
$[H,\vec{P}]=[H,\vec{P}\cdot\vec{P}]=0$. 
We will evaluate $\omega\langle\Phi_0|\vec{P}|\Phi_1\rangle$ and
$\omega\langle\Phi_0|\vec{P}\cdot\vec{P}|\Phi_2\rangle$,
where $|\Phi_1\rangle$ and $|\Phi_2\rangle$ are
the spurious states excited with
$\vec{P}$ and $\vec{P}\cdot\vec{P}$, respectively, and show that
these states have zero excitation
energy in STDDM. In the case of the exact problem, it is, with eqs.(\ref{var1}) and (\ref{var2}),
trivial to see
that $\omega\langle\Phi_0|\vec{P}|\Phi_1\rangle$ and
$\omega\langle\Phi_0|\vec{P}\cdot\vec{P}|\Phi_2\rangle$ are identical to zero because
the left-hand sides of eqs.(\ref{var1}) and (\ref{var2}) are reduced to the commutation relations
between the hamiltonian and these translational operators. Since the linearization and
the HF assumption are made in the derivation of STDDM, it is not trivial to show the
above properties of the spurious states. 
However, the linearization should be valid in the weak coupling limit and therefore
we can anticipate that the Goldstone theorem also holds in this case, provided that the
linearization procedure is correctly performed.

\subsection{Spurious states in RPA}
RPA gives zero excitation energy to the spurious
state $|\Phi_1\rangle$ excited with
$\vec{P}$, although only the 1p-1h and 1h-1p components of
the one-body amplitudes are taken into account.
To illustrate our approach for the problem of the spurious states,
we begin with proving
$\omega\langle\Phi_0|\vec{P}|\Phi_1\rangle=0$
in RPA.
Using the relation $\langle\Phi_0|a^+_{\alpha'}a_{\alpha}|\Phi_1\rangle
=x_{\alpha\alpha'}$ and eqs.(\ref{rpa1}) and (\ref{rpa2}) for $x_{ph}$ and $x_{hp}$,
we modify $\omega\langle\Phi_0|\vec{P}|\Phi_1\rangle$ as
\begin{eqnarray}
\omega\langle\Phi_0|i\vec{P}|\Phi_1\rangle&=&
\omega\sum_{ph}[\langle h|\nabla|p\rangle x_{ph}
+\langle p|\nabla|h\rangle x_{hp}] \nonumber \\
&=&\sum_{ph}
[\langle h|\nabla|p\rangle(\epsilon_{p}-\epsilon_{h})x_{ph}
+\langle p|\nabla|h\rangle(\epsilon_{h}-\epsilon_{p})x_{hp}] \nonumber \\
&+& 
\sum_{php'h'}[\langle h|\nabla|p\rangle(
\langle ph'|v|hp'\rangle_Ax_{p'h'}
+\langle pp'|v|hh'\rangle_Ax_{h'p'}) \nonumber \\
&-&\langle p|\nabla|h\rangle(
\langle hp'|v|ph'\rangle_Ax_{h'p'}
+\langle hh'|v|pp'\rangle_A)x_{p'h'}] .
\end{eqnarray}
A further modification is made 
using $h_0\psi_{\alpha}=\epsilon_{\alpha}\psi_{\alpha}$, where $h_0$ is the HF single-particle hamiltonian, and 
the closure relation
$\sum_{p}\psi_{p}(\vec{r})\psi_{p}^*(\vec{r'})
=\delta^3(\vec{r}-\vec{r'})-\sum_{h}\psi_{h}(\vec{r})\psi_{h}^*(\vec{r'})$:
\begin{eqnarray}
\omega\langle\Phi_0|i\vec{P}|\Phi_1\rangle
&=&\sum_{ph}
[\langle h|[\nabla,h_0]|p\rangle x_{ph}
+\langle h|[\nabla,h_0]|h\rangle x_{hp}] \nonumber \\
&+& 
\sum_{hph'}[(
\langle hh'|\nabla_1v|hp\rangle_A
-\langle hh'|v\nabla_1|hp\rangle
+\langle hh'|v\nabla_2|ph\rangle)x_{ph'}
\nonumber \\
&+&(\langle hp|\nabla_1v|hh'\rangle_A
-\langle hp|v\nabla_1|hh'\rangle
+\langle hp|v\nabla_2|h'h\rangle)x_{h'p})].
\label{sprpa1}
\end{eqnarray}
The first term on the right-hand side of the above equation
can be written in terms of $v$ using
\begin{eqnarray}
\langle\alpha'|[\nabla, h_0]|\alpha\rangle
&=&\sum_{h}[\langle\alpha'h|(\nabla_1v)|\alpha h\rangle_A 
\nonumber \\
&-&\langle\alpha'h|v\nabla_1|h\alpha\rangle
+\langle\alpha'h|v\nabla_2|h\alpha\rangle].
\label{sprpa2}
\end{eqnarray}
Finally eq.(\ref{sprpa1}) becomes
\begin{eqnarray}
\omega\langle\Phi_0|i\vec{P}|\Phi_1\rangle
&=&\sum_{phh'}[
(\langle hh'|(\nabla_1v)|ph'\rangle_A
-\langle hh'|v\nabla_1|h'p\rangle
+\langle hh'|v\nabla_2|h'p\rangle)x_{ph}
\nonumber \\
&+&(\langle ph'|(\nabla_1v)|hh'\rangle_A
-\langle ph'|v\nabla_1|h'h\rangle
+\langle ph'|v\nabla_2|h'h\rangle)x_{hp}]
\nonumber \\
&+& 
\sum_{phh'}[(
\langle h'h|(\nabla_1v)|h'p\rangle_A
-\langle h'h|v\nabla_1|ph'\rangle
+\langle h'h|v\nabla_2|ph'\rangle)x_{ph}
\nonumber \\
&+&(\langle h'p|(\nabla_1v)|h'h\rangle_A
-\langle h'p|v\nabla_1|hh'\rangle
+\langle h'p|v\nabla_2|hh'\rangle)x_{hp})],
\label{sprpa3}
\end{eqnarray}
where $(\nabla_1v)$ means that $\nabla_1$ acts only on $v$.
Since $v$ has translational invariance, the sum of the 
following two terms on the right-hand side of eq.(\ref{sprpa3}) becomes zero 
\begin{eqnarray}
\langle hh'|(\nabla_1v)|ph'\rangle_A
+\langle h'h|(\nabla_1v)|h'p\rangle_A
=\langle hh'|(\nabla_1v)+(\nabla_2v)|ph'\rangle_A=0.
\end{eqnarray}
Another sum of the two terms also vanishes
\begin{eqnarray}
-\langle hh'|v\nabla_1|h'p\rangle
+\langle h'h|v\nabla_2|ph'\rangle
=\langle h'h|-v\nabla_2+v\nabla_2|ph'\rangle=0.
\end{eqnarray}
Similarly, all other terms on the right-hand side of eq.(\ref{sprpa3}) cancel out. This means 
$\omega=0$.
As shown above, both the inclusion of the backward amplitude
$x_{hp}$ and the unrestricted sum over unoccupied single-particle
states are essential in RPA 
to give zero excitation energy to the spurious state.

\subsection{Spurious state in STDDM}
Along the lines illustrated above, we then show that
$\omega\langle\Phi_0|\vec{P}|\Phi_1\rangle=0$ in STDDM.
Using the equation for $x_{\alpha\alpha'}$ (eq.(\ref{stddm1})), 
we modify $\omega\langle\Phi_0|\vec{P}|\Phi_1\rangle$ as
\begin{eqnarray}
\omega\langle\Phi_0|i\vec{P}|\Phi_1\rangle
&=&
\sum_{\alpha\alpha'}\langle\alpha'|\nabla|\alpha\rangle\omega
x_{\alpha\alpha'} \nonumber \\
&=&\sum_{\alpha\alpha'}\langle\alpha'|\nabla|\alpha\rangle
\{(\epsilon_{\alpha}-\epsilon_{\alpha'})x_{\alpha\alpha'}
\nonumber \\
&+& 
(f_{\alpha'}-f_{\alpha})\sum_{\lambda\lambda'}\langle\alpha\lambda|v|\alpha'\lambda'\rangle_A
x_{\lambda'\lambda} \nonumber \\
&+& \sum_{\lambda\lambda'\lambda''}[X_{\lambda\lambda'\alpha'\lambda''}
\langle\alpha\lambda''|v|\lambda\lambda'\rangle
- X_{\alpha\lambda'\lambda\lambda''}
\langle\lambda\lambda''|v|\alpha'\lambda'\rangle]\},
\label{spstddm1}
\end{eqnarray}
where the sums are over both occupied and
unoccupied single-particle states.
Using $h_0\psi_{\alpha}=\epsilon_{\alpha}\psi_{\alpha}$ and 
the closure relation
$\sum_{\alpha}\psi(\vec{r})\psi_{\alpha}^*(\vec{r'})=\delta^3(\vec{r}-\vec{r'})$,
we further modify eq.(\ref{spstddm1}) as
\begin{eqnarray}
\omega\langle\Phi_0|i\vec{P}|\Phi_1\rangle&=&
\sum_{\alpha\alpha'}\langle\alpha'|[\nabla, h_0]|\alpha\rangle x_{\alpha\alpha'}
\nonumber \\
&+& 
\sum_{\lambda\lambda'h}[\langle h\lambda|\nabla_1v|h\lambda'\rangle_A
- (\langle h\lambda|v\nabla_1|h\lambda'\rangle -
\langle h\lambda|v\nabla_2|\lambda'h\rangle)]
x_{\lambda'\lambda} \nonumber \\
&+& \sum_{\alpha\alpha'\lambda\lambda'\lambda''}
[X_{\lambda\lambda'\alpha'\lambda''}
\langle\alpha'\lambda''|\nabla_1v|\lambda\lambda'\rangle
- X_{\alpha\lambda'\lambda\lambda''}
\langle\lambda\lambda''|v\nabla_1|\alpha\lambda'\rangle] .
\label{spstddm2}
\end{eqnarray}
Using eq.(\ref{sprpa2})
and changing summation indices, we obtain
\begin{eqnarray}
\omega\langle\Phi_0|i\vec{P}|\Phi_1\rangle&=&
\sum_{\alpha\alpha'h}[\langle\alpha'h|(\nabla_1v)|\alpha h\rangle_A 
-\langle\alpha' h|v\nabla_1|h\alpha\rangle
+\langle\alpha'h|v\nabla_2|h\alpha\rangle
\nonumber \\
&+& \langle h\alpha'|(\nabla_1v)|h\alpha\rangle_A
-\langle h\alpha'|v\nabla_1|\alpha h\rangle
+\langle h\alpha'|v\nabla_2|\alpha h\rangle)]
x_{\alpha\alpha'}
\nonumber \\
&+& \sum_{\alpha\beta\alpha'\beta'}X_{\alpha\beta\alpha'\beta'}
\langle\alpha'\beta'|(\nabla_1v)|\alpha\beta\rangle .
\label{spstddm3}
\end{eqnarray}
The first sum on the right-hand side of the above 
equation
which includes terms proportional to $x_{\alpha\alpha'}$ is a generalization
of eq.(\ref{sprpa3}) and vanishes for an interaction $v$ with translational invariance.
The second term on the right-hand side of eq.(\ref{spstddm3}) can be expressed as 
\begin{eqnarray}
\sum_{\alpha\beta\alpha'\beta'}X_{\alpha\beta\alpha'\beta'}
\langle\alpha'\beta'|(\nabla_1v)|\alpha\beta\rangle
\nonumber 
\end{eqnarray}
\begin{eqnarray}
&=&\frac{1}{2}
\sum_{\alpha\beta\alpha'\beta'}X_{\alpha\beta\alpha'\beta'}
(\langle\alpha'\beta'|(\nabla_1v)|\alpha\beta\rangle+
\langle\beta'\alpha'|(\nabla_1v)|\beta\alpha\rangle)
\nonumber \\
&=&\frac{1}{2}
\sum_{\alpha\beta\alpha'\beta'}X_{\alpha\beta\alpha'\beta'}
\langle\alpha'\beta'|(\nabla_1v)+(\nabla_2v)|\alpha\beta\rangle .
\label{spstddm4}
\end{eqnarray}
Since $v$ has translational invariance, $(\nabla_1v)+(\nabla_2v)=0$.
Thus again $\omega\langle\Phi_0|\vec{P}|\Phi_1\rangle=0$ is proven.
As shown above, unrestricted summation over single-particle indices
$\alpha$ and $\alpha'$ in eq.(\ref{spstddm1})
is essential to derive the last term on the right-hand side of eq.(\ref{spstddm2}). 
This means
that any ERPA formalisms with restricted one-body amplitudes
cannot give zero excitation energy to the spurious state 
associated with
the translational motion.
However, this does not depend on approximations for two-body amplitudes as long as
the symmetry property is respected as seen in eq.(\ref{spstddm4}).

\subsection{Double Spurious state in STDDM}
In a way similar to the above, we show that 
$\omega\langle\Phi_0|\vec{P}\cdot\vec{P}|\Phi_2\rangle=0$, where
$|\Phi_2\rangle$ is the double spurious state. 
The term
$\omega\langle\Phi_0|\vec{P}\cdot\vec{P}|\Phi_2\rangle$
contains both the one-body and two-body contributions,
\begin{eqnarray}
-\omega\langle\Phi_0|\vec{P}\cdot\vec{P}|\Phi_2\rangle&=&
\omega\{\sum_{\alpha\alpha'}(\langle\alpha'|\nabla^2|\alpha\rangle
-\sum_{h}2\langle\alpha'|\nabla|h\rangle\cdot
\langle h|\nabla|\alpha\rangle) x_{\alpha\alpha'}
\nonumber \\
&+&\sum_{\alpha\beta\alpha'\beta'}\langle\alpha'|\nabla|\alpha\rangle\cdot
\langle\beta'|\nabla|\beta\rangle X_{\alpha\beta\alpha'\beta'}\}.
\label{dspstddm1}
\end{eqnarray}
Using eqs.(\ref{stddm1}) and (\ref{stddm2}) for $x_{\alpha\alpha'}$ and 
$X_{\alpha\beta\alpha'\beta'}$, 
$h_0\psi_{\alpha}=\epsilon_{\alpha}\psi_{\alpha}$ and 
the closure relation
$\sum_{\alpha}\psi(\vec{r})\psi_{\alpha}^*(\vec{r'})=\delta^3(\vec{r}-\vec{r'})$,
we modify the right-hand side of the above equation. 
After some lengthy manipulations, the terms containing
$x_{\alpha\alpha'}$ and one summation
index over occupied single-particle states 
become 
\begin{eqnarray}
&2&\sum_{\alpha\alpha'h}[\langle\alpha'h|(\nabla^2_1v)|\alpha h\rangle_A
+\langle\alpha'h|(\nabla_1v)\cdot\nabla_1|\alpha h\rangle
\nonumber \\
&-&\langle\alpha'h|(\nabla_1v)\cdot\nabla_1|h\alpha\rangle
+\langle h\alpha'|(\nabla_1v)\cdot\nabla_1|h\alpha\rangle
-\langle h\alpha'|(\nabla_1v)\cdot\nabla_1|\alpha h\rangle]x_{\alpha\alpha'}
\nonumber \\
&+&2\sum_{\alpha\alpha'h}[\langle\alpha'h|(\nabla_1\cdot\nabla_2v)|\alpha h\rangle_A
+\langle\alpha'h|(\nabla_1v)\cdot\nabla_2|\alpha h\rangle
\nonumber \\
&-&\langle\alpha'h|(\nabla_1v)\cdot\nabla_2|h\alpha\rangle
+\langle\alpha'h|(\nabla_2v)\cdot\nabla_1|\alpha h\rangle
-\langle\alpha'h|(\nabla_2v)\cdot\nabla_1|h\alpha\rangle]x_{\alpha\alpha'},
\label{dspstddm2}
\end{eqnarray}
where the first sum comes from the terms with 
$x_{\alpha\alpha'}$ on the right-hand side of eq.(\ref{dspstddm1}) and
the second sum from the term with $X_{\alpha\beta\alpha'\beta'}$.
Since $v$ has translational invariance, 
$(\nabla^2_1v)+(\nabla_1\cdot\nabla_2v)=0$.
Therefore, the sum of the following two terms
in eq.(\ref{dspstddm2}) becomes $\langle\alpha'h|(\nabla^2_1v)|\alpha h\rangle_A+
\langle\alpha'h|(\nabla_1\cdot\nabla_2v)|\alpha h\rangle_A=0$.
All other terms vanish for similar reasons.
In addition to the terms shown in eq.(\ref{dspstddm2}),
there appear terms with $x_{\alpha\alpha'}$ and 
two summation indices over occupied 
single-particle states, 
and also terms with $X_{\alpha\beta\alpha'\beta'}$
in the modification process of eq.(\ref{dspstddm1}).
It is straightforward, though lengthy, to show that these terms
also vanish for a translationally invariant
interaction.
Thus $\omega\langle\Phi_0|\vec{P}\cdot\vec{P}|\Phi_2\rangle=0$,
that is, $\omega=0$.
As mentioned above unrestricted summation
over single-particle states is again essential
to obtain this conclusion. This means that only 
ERPA's with all one-body and two-body amplitudes,
that is, $x_{ph}, x_{hp}, x_{pp'}, x_{hh'}, X_{pp'hh'}, X_{hh'pp'},
X_{php'h'}, X_{phh'h''}, X_{h'h''ph}, X_{pp'hp''}$, $X_{hp''pp'}$,
$X_{pp'p''p'''}$ and $X_{hh'h''h'''}$,
give zero excitation energy
to the double-phonon state corresponding to
the spurious mode associated with the translational motion.

\subsection{Single and double Kohn modes}
When a system is confined to a harmonic potential $U=\frac{1}{2}m\omega_0^2r^2$, the
spurious mode associated with the translational motion has an eigenvalue of $\hbar\omega_0$,
independently of the translationally invariant two-body interaction. 
This property is known as the Kohn theorem \cite{Kohn,Brey,Dobson}.
In this subsection we show that our ERPA equations satisfy the Kohn theorem and also
that the eigenvalue of the double Kohn mode becomes $2\hbar\omega_0$. Due to the presence of the harmonic potential
the single-particle states are chosen to be eigenstates of the modified hamiltonian,
$h'\psi_{\alpha}=\epsilon_{\alpha}\psi_{\alpha}$, where $h'=h_0+\frac{1}{2}m\omega_0^2r^2$.
In a way similar to the spurious mode, we evaluate $\omega\langle\Phi_0|\vec{P}|\Phi_1\rangle$ using
the equations of motion in ERPA. Since the two-body interaction has translational invariance,
terms with the two-body interaction vanish and 
\begin{eqnarray}
\omega\langle\Phi_0|i\vec{P}|\Phi_1\rangle=-\sum_{\alpha\alpha'}
\langle\alpha|m\omega_0^2\vec{r}|\alpha'\rangle
x_{\alpha'\alpha}=-m\omega_0^2\langle\Phi_0|\vec{Q}|\Phi_1\rangle
\label{skohn1}
\end{eqnarray}
holds, where $\vec{Q}=\sum\langle\alpha|\vec{r}|\alpha'\rangle a^+_{\alpha}a_{\alpha'}$.
Similary, non-vanishing contribution to $\omega\langle\Phi_0|\vec{Q}|\Phi_1\rangle$ comes 
only from the kinetic energy term, and we obtain
\begin{eqnarray}
\omega\langle\Phi_0|\vec{Q}|\Phi_1\rangle=-\frac{\hbar^2}{m}\langle\Phi_0|i\vec{P}|\Phi_1\rangle.
\label{skohn2}
\end{eqnarray}
It is essential to keep all components of the one-body amplitudes to obtain the above expressions. 
From eqs.(\ref{skohn1}) and (\ref{skohn2}), we get $\omega=\pm\hbar\omega_0$. 

In the case of the double Kohn mode, expectation values of three operators couple in the following way,
\begin{eqnarray}
\omega\langle\Phi_0|i\vec{P}\cdot i\vec{P}|\Phi_2\rangle&=&
2m\omega_0^2\langle\Phi_0|\vec{Q}\cdot i\vec{P}|\Phi_2\rangle, 
\label{dkohn1}
\\
\omega\langle\Phi_0|\vec{Q}\cdot i\vec{P}|\Phi_2\rangle&=&
\frac{\hbar^2}{m}\langle\Phi_0|i\vec{P}\cdot i\vec{P}|\Phi_2\rangle+
m\omega_0^2\langle\Phi_0|\vec{Q}\cdot\vec{Q}|\Phi_2\rangle, 
\label{dkohn2}
\\
\omega\langle\Phi_0|\vec{Q}\cdot\vec{Q}|\Phi_2\rangle&=&
2\frac{\hbar^2}{m}\langle\Phi_0|\vec{Q}\cdot i\vec{P}|\Phi_2\rangle.
\label{dkohn3}
\end{eqnarray}
The right-hand side of eq.(\ref{dkohn1}) comes from the harmonic potential. Both the kinetic energy term 
and the harmonic potential contribute to the right-hand side of eq.(\ref{dkohn2}), and the kinetic energy term becomes
non-vanishing on the right-hand side of eq.(\ref{dkohn3}). All terms with the two-body interaction vanish due to translational
invariance. 
It is essential to keep all components of
the one-body and two-body amplitudes to derive eqs.(\ref{dkohn1})-(\ref{dkohn3}) as in the case of the double spurious mode
discussed in Subsect. 3.4.
From the above equations we get $\omega=\pm2\hbar\omega_0$.

\subsection{Continuity equations}
We end this section by showing that our ERPA equations satisfy continuity equations.
In a way similar to the single spurious mode 
we evaluate $\omega_{\lambda}\langle\Phi_0|\hat{\rho}(\vec{r})|\Phi_{\lambda}\rangle$,
where $\hat{\rho}$ is the density operator $\hat{\rho}(\vec{r})=
\sum\psi^*_{\alpha}(\vec{r})\psi_{\alpha'}(\vec{r})a^+_{\alpha}a_{\alpha'}$, and 
obtain
\begin{eqnarray}
\omega_{\lambda}\langle\Phi_0|\hat{\rho}(\vec{r})|\Phi_{\lambda}\rangle
=-\nabla\cdot\langle\Phi_0|\vec{j}(\vec{r})|\Phi_{\lambda}\rangle,
\label{CE1}
\end{eqnarray}
where the current operator $\vec{j}$ is given by 
\begin{eqnarray}
\vec{j}(\vec{r})=\frac{\hbar^2}{2m}\sum[\psi^*_{\alpha}(\vec{r})\nabla\psi_{\alpha'}(\vec{r})
-(\nabla\psi^*_{\alpha}(\vec{r}))\psi_{\alpha'}(\vec{r})
]
a^+_{\alpha}a_{\alpha'}
\end{eqnarray}
for a momentum-independent two-body interaction. 
Thus the continuity equation for the one-body transition density
and current is satisfied. Keeping all components of the one-body amplitudes is
essential to obtain the continuity equation.

Similarly, the transition amplitude for the two-body density operator $\hat{\rho_2}(\vec{r},\vec{r}')$
defined by 
\begin{eqnarray}
\hat{\rho_2}(\vec{r},\vec{r}')=\sum_{\alpha\beta\alpha'\beta'}\psi^*_{\alpha}(\vec{r})\psi^*_{\beta}(\vec{r}')\psi_{\beta'}(\vec{r}')\psi_{\alpha'}(\vec{r})
a^+_{\alpha}a^+_{\beta}a_{\beta'}a_{\alpha'}
\end{eqnarray}
satisfies the continuity equation
\begin{eqnarray}
\omega_{\lambda}\langle\Phi_0|\hat{\rho_2}(\vec{r},\vec{r}')|\Phi_{\lambda}\rangle
=-(\nabla_r\cdot\langle\Phi_0|\vec{j}_2(\vec{r},\vec{r}')|\Phi_{\lambda}\rangle+
\nabla_{r'}\cdot\langle\Phi_0|\vec{j}_2(\vec{r}',\vec{r})|\Phi_{\lambda}\rangle),
\label{CE2}
\end{eqnarray}
where the two-body current operator $\vec{j}_2$ for a momentum-independent two-body interaction is given by
\begin{eqnarray}
\vec{j}_2(\vec{r},\vec{r}')&=&\frac{\hbar^2}{2m}
\sum[\psi^*_{\alpha}(\vec{r})(\nabla\psi_{\alpha'}(\vec{r}))
-(\nabla\psi^*_{\alpha}(\vec{r}))\psi_{\alpha'}(\vec{r}))]
\psi^*_{\beta}(\vec{r}')\psi_{\beta'}(\vec{r}')
a^+_{\alpha}a^+_{\beta}a_{\beta'}a_{\alpha'}.
\end{eqnarray}
In the derivation of eq.(\ref{CE2}) it is again essential to keep all components of the one-body
and two-body amplitudes.
\section{ERPA with hermiticity}
The equations of STDDM (eqs.(\ref{stddm1}) and (\ref{stddm2})) show
asymmetry and non-hermiticity, although this causes no problem in conserving various physical properties as discussed
above. 
In the following we show that
ERPA with symmetry and hermiticity can be formulated using
the equation-of-motion approach \cite{Rowe} and
the correlated ground state in TDDM.
We have pointed out \cite{T90}, in deriving the Landau's expression
for the spreading width of a collective state,
that it is important to include ground-state correlations to remove the asymmetry in STDDM.
It is well-known \cite{Rowe} that the
asymmetry problem always appears in the equation-of-motion approach when the ground state is replaced by an approximate one.
Before presenting the formulation of our ERPA, therefore,
we summarize the origin of the asymmetry in the equation-of-motion approach.
When $|\Phi_0\rangle$ is the exact ground state of the hamiltonian, there 
exists an identity involving a one-body operator $A=a^+_{\alpha}a_{\alpha'}$ 
and a two-body operator $B=a^+_{\alpha}a^+_{\beta}a_{\beta'}a_{\alpha'}$:
\begin{eqnarray}
\langle\Phi_0|[[B,H],A]|\Phi_0\rangle -
\langle\Phi_0|[[A,H],B]|\Phi_0\rangle
=\langle\Phi_0|[H,[A,B]]|\Phi_0\rangle =0.
\label{ident}
\end{eqnarray}
When $|\Phi_0\rangle$
is approximated
by the HF ground state, the above identity is violated, that is,
\begin{eqnarray}
\langle\Phi_0|[H, [A,B]]|\Phi_0\rangle
\neq0
\end{eqnarray}
and, consequently,
\begin{eqnarray}
\langle\Phi_0|[
[B,H],A]|\Phi_0\rangle
\neq
\langle\Phi_0|
[A,H],B]|\Phi_0\rangle.
\label{asym}
\end{eqnarray}
Since the left-hand side of the above equation describes the coupling of the
one-body amplitudes to the two-body ones, 
and the right-hand side,
that of the two-body amplitudes to the one-body ones,
the resulting ERPA has asymmetric couplings.
In order to avoid the difficulty of eq.(\ref{asym}),
Rowe introduced a symmetrized double commutator \cite{Rowe}. 
However, it was pointed out \cite{Adachi}
that there is an ambiguity in the choice of such a double commutator.

Now we proceed to the presentation of our ERPA with ground-state correlations.
The ground state $|\Phi_0\rangle$ in TDDM is constructed so that
\begin{eqnarray}
\langle\Phi_0|[H,a^+_{\alpha}a_{\alpha'}]|\Phi_0\rangle =0
\label{grc1}
\end{eqnarray}
and
\begin{eqnarray}
\langle\Phi_0|[H,a^+_{\alpha}a^+_{\beta}a_{\beta'}a_{\alpha'}]|\Phi_0\rangle =0
\label{grc2}
\end{eqnarray}
are satisfied for any single-particle indices \cite{TG1}. 
In other words 
the occupation matrix $n^0_{\alpha\alpha'}$ and the correlation matrix
$C^0_{\alpha\beta\alpha'\beta'}$, the expansion coefficients of $\rho_0$ and
$C_0$, respectively,
are determined in TDDM so that the above two equations are satisfied.
The explicit expression for eqs.(\ref{grc1}) and (\ref{grc2}) depends on the single-particle state $\psi_{\alpha}$.
The equations for $n^0_{\alpha\alpha'}$ and 
$C^0_{\alpha'\beta'\alpha\beta}$ shown
in Appendix A are obtained 
when $\psi_{\alpha}$ is chosen to be an eigenstate of the mean field hamiltonian $h_0(\rho_0)$,
that is, 
\begin{eqnarray}
h_0(\rho_0)\psi_{\alpha}(1)=-\frac{\hbar^2\nabla^2}{2m}\psi_{\alpha}(1)+\int d2 v(1,2)
[\rho_0(2,2)\psi_{\alpha}(1)-\rho_0(1,2)\psi_{\alpha}(2)]=\epsilon_{\alpha}\psi_{\alpha}(1),
\label{hf}
\end{eqnarray}
where
\begin{eqnarray}
\rho_0(11')&=&\sum_{\alpha\alpha'}n^0_{\alpha\alpha'}\psi_{\alpha}(1)
\psi_{\alpha'}^{*}(1').
\end{eqnarray}
Although it is not evident to find an
analytic solution of eqs.(\ref{grc1}) and (\ref{grc2}) \cite{TSW},
a method for obtaining $n^0_{\alpha\alpha'}$
and $C^0_{\alpha\beta\alpha'\beta'}$ numerically
has been proposed \cite{T95} and already been tested for realistic nuclei in the study of giant resonances 
built on the correlated ground state \cite{T98,T01}.
Since the commutation relation $[A,B]=[a^+_{\alpha}a_{\alpha'},
a^+_{\beta}a^+_{\gamma}a_{\gamma'}a_{\beta'}]$ in eq.(\ref{ident})
becomes a sum of two-body operators, we find
\begin{eqnarray}
\langle\Phi_0|[H, [a^+_{\alpha}a_{\alpha'},
a^+_{\beta}a^+_{\gamma}a_{\gamma'}a_{\beta'}]]|\Phi_0\rangle
=0
\end{eqnarray}
which
holds due to eq.(\ref{grc2}). This means that the coupling matrices are symmetric, that is,
\begin{eqnarray}
\langle\Phi_0|[
[a^+_{\beta}a^+_{\gamma}a_{\gamma'}a_{\beta'},H],a^+_{\alpha}a_{\alpha'}]|\Phi_0\rangle
=
\langle\Phi_0|[
[a^+_{\alpha}a_{\alpha'},H],a^+_{\beta}a^+_{\gamma}a_{\gamma'}a_{\beta'}]|\Phi_0\rangle
\end{eqnarray}
for the correlated ground state in TDDM.
The ERPA equations based on the TDDM 
ground state are formulated using the equation of motion
approach \cite{Rowe} as 
\begin{eqnarray}
\langle\Psi_0|[[a^+_{\alpha}a_{\alpha'},H],Q^+]|\Psi_0\rangle
=
\omega\langle\Psi_0|[a^+_{\alpha}a_{\alpha'},Q^+]|\Psi_0\rangle
\label{erpa1}
\end{eqnarray}
\begin{eqnarray}
\langle\Psi_0|[[:a^+_{\alpha}a^+_{\beta}a_{\beta'}a_{\alpha'}:,H],Q^+]|\Psi_0\rangle
=
\omega\langle\Psi_0|[:a^+_{\alpha}a^+_{\beta}a_{\beta'}a_{\alpha'}:,Q^+]|\Psi_0\rangle,
\label{erpa2}
\end{eqnarray}
where the operator $Q^+$ is defined by
\begin{eqnarray}
\hspace{0.5cm}Q^+=\sum(x_{\lambda\lambda'}a^+_{\lambda}a_{\lambda'}
+X_{\lambda_1\lambda_2\lambda_1'\lambda_2'}:a^+_{\lambda_1}a^+_{\lambda_2}a_{\lambda_2'}a_{\lambda_1'}:)
\label{q-opr}
\end{eqnarray}
and $|\Psi_0\rangle$ is assumed to have the following properties
\begin{eqnarray}
Q^+|\Psi_0\rangle&=&|\Psi\rangle\\
Q|\Psi_0\rangle&=&0.
\end{eqnarray} 
In eqs.(\ref{erpa2}) and (\ref{q-opr}), $:~:$ stands for 
$:a^+_{\alpha}a^+_{\beta}a_{\beta'}a_{\alpha'}:=
a^+_{\alpha}a^+_{\beta}a_{\beta'}a_{\alpha'}-{\cal A}(a^+_{\alpha}a_{\alpha'}
\langle\Psi_0|a^+_{\beta}a_{\beta'}|\Psi_0\rangle+a^+_{\beta}a_{\beta'}
\langle\Psi_0|a^+_{\alpha}a_{\alpha'}|\Psi_0\rangle)$,
where ${\cal A}$ is an antisymmetrization operator.
The above equation can be written in matrix form
\begin{eqnarray}
\left(
\begin{array}{cc}
A & C \\
B & D
\end{array}
\right)
\left(
\begin{array}{c}
x \\
X
\end{array}
\right)
=\omega
\left(
\begin{array}{cc}
S_1 & T_1 \\
T_2 & S_2
\end{array}
\right)
\left(
\begin{array}{c}
x \\
X
\end{array}
\right) ,
\label{erpa3}
\end{eqnarray}
where each matrix element is given by
\begin{eqnarray}
S_1(\alpha'\alpha:\lambda\lambda')
&=&\langle\Psi_0|[a^+_{\alpha}a_{\alpha'},a^+_{\lambda}a_{\lambda'}]|\Psi_0\rangle,
\label{S1}
\\
S_2(\alpha'\beta'\alpha\beta:\lambda_1\lambda_2\lambda_1'\lambda_2')
&=&\langle\Psi_0|[
:a^+_{\alpha}a^+_{\beta}a_{\beta'}a_{\alpha'}:,:a^+_{\lambda_1}a^+_{\lambda_2}a_{\lambda_2'}a_{\lambda_1'}:]|\Psi_0\rangle,
\\
T_1(\alpha'\alpha:\lambda_1\lambda_2\lambda_1'\lambda_2')
&=&\langle\Psi_0|[
a^+_{\alpha}a_{\alpha'},:a^+_{\lambda_1}a^+_{\lambda_2}a_{\lambda_2'}a_{\lambda_1'}:]|\Psi_0\rangle ,
\\
T_2(\alpha'\beta'\alpha\beta:\lambda\lambda')
&=&\langle\Psi_0|[
:a^+_{\alpha}a^+_{\beta}a_{\beta'}a_{\alpha'}:,a^+_{\lambda}a_{\lambda'}]|\Psi_0\rangle,
\label{T2}
\\
A(\alpha'\alpha:\lambda\lambda')&=&\langle\Psi_0|[
[a^+_{\alpha}a_{\alpha'},H],a^+_{\lambda}a_{\lambda'}]|\Psi_0\rangle,
\\
B(\alpha'\beta'\alpha\beta:\lambda\lambda')
&=&\langle\Psi_0|[
[:a^+_{\alpha}a^+_{\beta}a_{\beta'}a_{\alpha'}:,H],a^+_{\lambda}a_{\lambda'}]|\Psi_0\rangle,
\\
C(\alpha'\alpha:\lambda_1\lambda_2\lambda_1'\lambda_2')
&=&\langle\Psi_0|[
[a^+_{\alpha}a_{\alpha'},H],:a^+_{\lambda_1}a^+_{\lambda_2}a_{\lambda_2'}a_{\lambda_1'}:]|\Psi_0\rangle,
\\
D(\alpha'\beta'\alpha\beta:\lambda_1\lambda_2\lambda_1'\lambda_2')
&=&\langle\Psi_0|[
[:a^+_{\alpha}a^+_{\beta}a_{\beta'}a_{\alpha'}:,H],:a^+_{\lambda_1}a^+_{\lambda_2}a_{\lambda_2'}a_{\lambda_1'}:]|\Psi_0\rangle .
\end{eqnarray}
When the above matrices are evaluated,
the ground state $|\Psi_0\rangle$ is replaced by $|\Phi_0\rangle$ in TDDM.
Then all matrices in the above are written in terms of
$n^0_{\alpha\alpha'}$
and $C^0_{\alpha\beta\alpha'\beta'}$, which are shown in Appendix B. 
Due to eqs.(\ref{grc1}) and (\ref{grc2}), the above matrices have the following symmetries
\begin{eqnarray}
A(\alpha'\alpha:\lambda\lambda')&=&A(\lambda'\lambda:\alpha\alpha')=A(\lambda\lambda':\alpha'\alpha)^*,
\\
B(\alpha'\beta'\alpha\beta:\lambda\lambda')&=&C(\lambda'\lambda:\alpha\beta\alpha'\beta')
=C(\lambda\lambda':\alpha'\beta'\alpha\beta)^* .
\end{eqnarray}
This version of ERPA gives zero excitation energy to spurious modes associated with
operators ${\cal O}$ which commute with $H$ and consist of 
one-body and (or) two-body operators. This is because $\omega\langle\Psi_0|{\cal O}|\Psi\rangle
=\langle\Psi_0|[H,{\cal O}]|\Psi\rangle=0$ holds due to eqs.(\ref{erpa1}) and (\ref{erpa2}).
Although the coupling matrix between the one-body and two-body amplitudes is symmetric, 
the hamiltonian matrix on the left-hand side of eq.(\ref{erpa3}) is not yet hermitian because $D=D^+$ does not hold. 
This originates in the fact that
$\langle\Phi_0|
[H,[:a^+_{\alpha}a^+_{\beta}a_{\beta'}a_{\alpha'}:,
:a^+_{\lambda_1}a^+_{\lambda_2}a_{\lambda_2'}a_{\lambda_1'}:]]|\Phi_0\rangle\neq 0$.
In order to obtain a hermitian hamiltonian matrix without any truncation of the two-body amplitudes, we
need to impose
\begin{eqnarray}
\langle\Phi_0|[H,a^+_{\alpha}a^+_{\beta}a^+_{\gamma}a_{\gamma'}a_{\beta'}a_{\alpha'}]|\Phi_0\rangle =0
\label{grc3}
\end{eqnarray}
in addition to eqs.(\ref{grc1}) and (\ref{grc2}). This condition guarantees 
$\langle\Phi_0|
[H,[:a^+_{\alpha}a^+_{\beta}a_{\beta'}a_{\alpha'}:,
:a^+_{\lambda_1}a^+_{\lambda_2}a_{\lambda_2'}a_{\lambda_1'}:]]|\Phi_0\rangle=0$, 
and thereby 
\begin{eqnarray}
D(\alpha'\beta'\alpha\beta:\lambda_1\lambda_2\lambda_1'\lambda_2')
=D(\lambda_1'\lambda_2'\lambda_1\lambda_2:\alpha\beta\alpha'\beta')
=D(\lambda_1\lambda_2\lambda_1'\lambda_2':\alpha'\beta'\alpha\beta)^*.
\end{eqnarray}
Equation (\ref{grc3}) is explicitly shown in Appendix B.
For a hermitian hamiltonian matrix
the ortho-normal condition is given by \cite{Taka}
\begin{eqnarray}
(x^*_{\mu'} X^*_{\mu'})\left(
\begin{array}{cc}
S_1 & T_1 \\
T_2& S_2
\end{array}
\right)
\left(
\begin{array}{c}
x_{\mu} \\
X_{\mu}
\end{array}
\right)
=\delta_{\mu\mu'},
\end{eqnarray}
where $x_{\mu}$ and $X_{\mu}$ constitute an eigenstate of eq.(\ref{erpa3}) with $\omega=\omega_{\mu}$.
The completeness relation becomes
\begin{eqnarray}
\sum_{\mu}
\left(
\begin{array}{c}
x_{\mu} \\
X_{\mu}
\end{array}
\right)
(x^*_{\mu} X^*_{\mu})
\left(
\begin{array}{cc}
S_1 & T_1 \\
T_2& S_2
\end{array}
\right)
=I .
\end{eqnarray} 
The transition amplitudes for one-body and two-body operators,
$z=\langle\Psi_0|a^+_{\alpha}a_{\alpha'}|\Psi\rangle$ and 
$Z=\langle\Psi_0|:a^+_{\alpha}a^+_{\beta}a_{\beta'}a_{\alpha'}:|\Psi\rangle$, respectively,
are calculated as follows
\begin{eqnarray}
\left(
\begin{array}{c}
z \\
Z
\end{array}
\right)=
\left(
\begin{array}{cc}
S_1 & T_1 \\
T_2& S_2
\end{array}
\right)
\left(
\begin{array}{c}
x \\
X
\end{array}
\right).
\end{eqnarray} 
Equation (\ref{erpa3}) has a certain similarity with the so-called 
Self-Consistent RPA (SCRPA) equations \cite{Rowe,SCRPA1,SCRPA2}, extended to include
higer configurations.
In case the $X_{\lambda_1\lambda_2\lambda_1'\lambda_2'}$ amplitudes are dropped
in eq.(\ref{q-opr}), eq.(\ref{erpa3}) reduces to something similar to 
what has become known as renormalized RPA (r-RPA) \cite{r-RPA1}. 
The main difference seems to come from the fact that here eq.(\ref{grc1})
serves to determine the occupation matrix $n^0_{\alpha\alpha'}$ whereas
in r-RPA eq.(\ref{grc1}) is used to establish the single particle basis.
It should be interesting to investigate this relation more in detail in the future.

\section{Summary}
Necessary conditions that the spurious state associated with
the translational motion and its double-phonon state have zero
excitation energy in extended ERPA (ERPA) were investigated
using the small amplitude limit of the time-dependent density-matrix theory (STDDM).
The reason why STDDM was used
is that it has a quite general form of the ERPA kind
based on the HF ground state.
In the case of the single spurious state it is found that ERPA which keeps all components
of the one-body amplitudes 
gives the spurious state at zero excitation energy. This does not depend on approximations
for the two-body amplitudes as long as they are properly antisymmetrized. For example, ERPA
with only the 2p-2h and 2h-2p components of the two-body amplitudes preserves this property
of the single spurious state.
In the case of the double spurious state,
all components of the two-body amplitudes are found necessary to yield the mode at zero excitation energy.
Of course, no truncation in 
single-particle space should be made in both cases. 
The Kohn theorem for the single and double Kohn modes and the continuity equations for transition
densities and currents were also investigated and found to hold under the same conditions as those
necessary for the spurious states. 
It was pointed out that STDDM inherently has asymmetry and non-hermiticity,
although it conserves various physical properties as mentioned above.
A formulation of ERPA with hermiticity was also presented using TDDM, in which it was discussed
that a three-body correlation matrix needs to be included in the description of ground-state correlations.
The investigations in this work were performed for the spurious translational motion. It seems, however,
clear that analogous considerations can be made for any spontaneously broken symmetry.
An interesting case could be the coupling of quark-antiquark to the four quark sector using, e.g. 
the Nambu-Jona-Lasinio model \cite{Nambu}. In this case one knows that chiral symmetry is spontaneously
broken \cite{Nambu} and therefore in the chiral limit a double Goldstone mode (two pions) should appear.
In the case of finite current quark masses analogous equations to those yielding the Kohn modes considered
here should exist, actually well known as the Gellmann-Oakes-Renner relation \cite{Gell}.

\appendix
\section{}
When $\psi_{\alpha}$ is chosen to be an eigenstate of the mean field hamiltonian (eq.(\ref{hf})),
eqs.(\ref{grc1}) and (\ref{grc2}) become
\begin{eqnarray}
(\epsilon_{\alpha'}-\epsilon_{\alpha})n^0_{\alpha\alpha'}&=&
\sum_{\lambda_1\lambda_2\lambda_3}(C^0_{\lambda_1\lambda_2\alpha'\lambda_3}
\langle\alpha\lambda_3|v|\lambda_1\lambda_2\rangle -C^0_{\alpha\lambda_3\lambda_1\lambda_2}
\langle\lambda_1\lambda_2|v|\alpha'\lambda_3\rangle)
\label{Agrc1}
\\
(\epsilon_{\alpha'}+\epsilon_{\beta'}-\epsilon_{\alpha}-\epsilon_{\beta})
C^0_{\alpha\beta\alpha'\beta'}&=&
B^0_{\alpha\beta\alpha'\beta'}+P^0_{\alpha\beta\alpha'\beta'}+H^0_{\alpha\beta\alpha'\beta'},
\label{Agrc2}
\end{eqnarray}
where
\begin{eqnarray}
B^0_{\alpha\beta\alpha'\beta'}&=&\sum_{\lambda_1\lambda_2\lambda_3\lambda_4}
\langle\lambda_1\lambda_2|v|\lambda_3\lambda_4\rangle_A
[(\delta_{\alpha\lambda_1}-n^0_{\alpha\lambda_1})(\delta_{\beta\lambda_2}-n^0_{\beta\lambda_2})
n^0_{\lambda_3\alpha'}n^0_{\lambda_4\beta'}
\nonumber \\
&-&n^0_{\alpha\lambda_1}n^0_{\beta\lambda_2}(\delta_{\lambda_3\alpha'}-n^0_{\lambda_3\alpha'})
(\delta_{\lambda_4\beta'}n^0_{\lambda_4\beta'})],
\\
P^0_{\alpha\beta\alpha'\beta'}&=&\sum_{\lambda_1\lambda_2\lambda_3\lambda_4}
\langle\lambda_1\lambda_2|v|\lambda_3\lambda_4\rangle
[(\delta_{\alpha\lambda_1}\delta_{\beta\lambda_2}
-\delta_{\alpha\lambda_1}n^0_{\beta\lambda_2}
-n^0_{\alpha\lambda_1}\delta_{\beta\lambda_2})
C^0_{\lambda_3\lambda_4\alpha'\beta'}
\nonumber \\
&-&(\delta_{\lambda_3\alpha'}\delta_{\lambda_4\beta'}
-\delta_{\lambda_3\alpha'}n^0_{\lambda_4\beta'}
-n^0_{\lambda_3\alpha'}\delta_{\lambda_4\beta'})
C^0_{\alpha\beta\lambda_1\lambda_2}],
\\
H^0_{\alpha\beta\alpha'\beta'}&=&\sum_{\lambda_1\lambda_2\lambda_3\lambda_4}
\langle\lambda_1\lambda_2|v|\lambda_3\lambda_4\rangle_A
[\delta_{\alpha\lambda_1}(n^0_{\lambda_3\alpha'}C^0_{\lambda_4\beta\lambda_2\beta'}
-n^0_{\lambda_3\beta'}C^0_{\lambda_4\beta\lambda_2\alpha'}+C^0_{\lambda_3\lambda_4\beta\alpha'\lambda_2\beta'})
\nonumber \\
&+&\delta_{\beta\lambda_2}(n^0_{\lambda_4\beta'}C^0_{\lambda_3\alpha\lambda_1\alpha'}
-n^0_{\lambda_4\alpha'}C^0_{\lambda_3\alpha\lambda_2\beta'}+C^0_{\lambda_4\lambda_3\alpha\beta'\lambda_2\alpha'})
\nonumber \\
&-&\delta_{\alpha'\lambda_3}(n^0_{\alpha\lambda_1}C^0_{\lambda_4\beta\lambda_2\beta'}
-n^0_{\beta\lambda_1}C^0_{\lambda_4\alpha\lambda_2\beta'}+C^0_{\alpha\lambda_4\beta\lambda_1\lambda_2\beta'})
\nonumber \\
&-&\delta_{\beta'\lambda_4}(n^0_{\beta\lambda_2}C^0_{\lambda_3\alpha\lambda_1\alpha'}
-n^0_{\alpha\lambda_2}C^0_{\lambda_3\beta\lambda_1\alpha'}+C^0_{\beta\lambda_3\alpha\lambda_2\lambda_1\alpha'})].
\label{H0}
\end{eqnarray}
The three-body correlation matrix $C^0_{\alpha\beta\gamma\alpha'\beta'\gamma'}$ is also included in eq.(\ref{H0}).
The equation for $C^0_{\alpha\beta\gamma\alpha'\beta'\gamma'}$ is obtained
by neglecting four-body amplitudes and becomes
\begin{eqnarray}
(\epsilon_{\alpha'}+\epsilon_{\beta'}+\epsilon_{\gamma'}-\epsilon_{\alpha}-\epsilon_{\beta}-\epsilon_{\gamma})
C^0_{\alpha\beta\gamma\alpha'\beta'\gamma'}&=&U_{\alpha\beta\gamma\alpha'\beta'\gamma'}
+U_{\alpha\beta\gamma\beta'\gamma'\alpha'}-U_{\alpha\beta\gamma\alpha'\gamma'\beta'}
\nonumber \\
&-&U^*_{\alpha'\beta'\gamma'\alpha\beta\gamma}-U^*_{\alpha'\beta'\gamma'\beta\gamma\alpha}
+U^*_{\alpha'\beta'\gamma'\alpha\gamma\beta}
\nonumber \\
&+&V_{\alpha\beta\gamma\alpha'\beta'\gamma'}
+V_{\alpha\beta\gamma\beta'\gamma'\alpha'}-V_{\alpha\beta\gamma\alpha'\gamma'\beta'}
\nonumber \\
&-&V^*_{\alpha'\beta'\gamma'\alpha\beta\gamma}-V^*_{\alpha'\beta'\gamma'\beta\gamma\alpha}
+V^*_{\alpha'\beta'\gamma'\alpha\gamma\beta},
\label{Agrc3}
\end{eqnarray}
where
\begin{eqnarray}
U_{\alpha\beta\gamma\alpha'\beta'\gamma'}&=-&\sum_{\lambda_1\lambda_2}[
\langle\lambda_1\lambda_2|v|\alpha'\beta'\rangle_A(n^0_{\gamma\lambda_1}C^0_{\alpha\beta\lambda_2\gamma'}
-n^0_{\beta\lambda_1}C^0_{\alpha\gamma\lambda_2\gamma'}+n^0_{\alpha\lambda_1}C^0_{\beta\gamma\lambda_2\gamma'})
\nonumber \\
&+&\langle\lambda_1\lambda_2|v|\alpha'\beta'\rangle C^0_{\alpha\beta\gamma\lambda_1\lambda_2\gamma'}],
\\
V_{\alpha\beta\gamma\alpha'\beta'\gamma'}&=-&\sum_{\lambda_1\lambda_2\lambda_3}
\langle\lambda_1\lambda_2|v|\alpha'\lambda_3\rangle(-n^0_{\lambda_3\gamma'}n^0_{\gamma\lambda_2}C^0_{\alpha\beta\lambda_1\beta'}
+n^0_{\lambda_3\gamma'}C^0_{\alpha\beta\gamma\lambda_1\lambda_2\beta'}
\nonumber \\
&+&C^0_{\alpha\beta\lambda_1\lambda_2}C^0_{\gamma\lambda_3\beta'\gamma'}
-C^0_{\alpha\beta\lambda_1\beta'}C^0_{\gamma\lambda_3\lambda_2\gamma'}
+{\rm all~other~exchange~terms }).
\end{eqnarray}
Equations for correlation matrices of higher ranks may be formulated according to the truncation rules given in
ref.\cite{Niita}.

\section{}
The matrix elements of eqs.(\ref{S1})-(\ref{T2}) are explicitly shown below.
\begin{eqnarray}
S_1(\alpha'\alpha:\lambda\lambda')
&=&n^0_{\lambda'\alpha}\delta_{\alpha'\lambda}-n^0_{\alpha'\lambda}\delta_{\alpha\lambda'},
\label{BS1} \\
T_1(\alpha'\alpha:\lambda_1\lambda_2\lambda_1'\lambda_2')
&=&C^0_{\lambda_1'\lambda_2'\alpha\lambda_2}\delta_{\alpha'\lambda_1}
+C^0_{\alpha'\lambda_1'\lambda_1\lambda_2}\delta_{\alpha\lambda_2'}
-C^0_{\alpha'\lambda_2'\lambda_1\lambda_2}\delta_{\alpha\lambda_1'}
-C^0_{\lambda_1'\lambda_2'\alpha\lambda_1}\delta_{\alpha'\lambda_2},
\label{BT1}
\\
T_2(\alpha'\beta'\alpha\beta:\lambda\lambda')
&=&C^0_{\lambda'\beta'\alpha\beta}\delta_{\alpha'\lambda}+C^0_{\alpha'\beta'\lambda\alpha}\delta_{\beta\lambda'}
-C^0_{\alpha'\beta'\lambda\beta}\delta_{\alpha\lambda'}-C^0_{\lambda'\alpha'\alpha\beta}\delta_{\beta'\lambda}
\label{BT2}\\
S_2(\alpha'\beta'\alpha\beta:\lambda_1\lambda_2\lambda_1'\lambda_2')
&=& {\cal A}(\delta_{\alpha'\lambda_1}\delta_{\beta'\lambda_2})({\cal A}(n^0_{\lambda_1'\alpha}n^0_{\lambda_2'\beta})
+C^0_{\lambda_1'\lambda_2'\alpha\beta}) 
-{\cal A}(\delta_{\alpha\lambda_1'}\delta_{\beta\lambda_2'})({\cal A}(n^0_{\alpha'\lambda_1}n^0_{\beta'\lambda_2})
+C^0_{\alpha'\beta'\lambda_1\lambda_2})
\nonumber \\
&+&F(\alpha'\beta'\alpha\beta:\lambda_1\lambda_2\lambda_1'\lambda_2')
-F(\alpha'\beta'\alpha\beta:\lambda_1\lambda_2\lambda_2'\lambda_1')
\nonumber \\
&-&F(\alpha'\beta'\beta\alpha:\lambda_1\lambda_2\lambda_1'\lambda_2')
+F(\alpha'\beta'\beta\alpha:\lambda_1\lambda_2\lambda_2'\lambda_1')
\nonumber \\
&-&F(\lambda_1'\lambda_2'\lambda_1\lambda_2:\alpha\beta\alpha'\beta')
+F(\lambda_1'\lambda_2'\lambda_2\lambda_1:\alpha\beta\alpha'\beta')
\nonumber \\
&+&F(\lambda_1'\lambda_2'\lambda_1\lambda_2:\alpha\beta\beta'\alpha')
-F(\lambda_1'\lambda_2'\lambda_2\lambda_1:\alpha\beta\beta'\alpha'),
\label{BS2}
\end{eqnarray}
where
\begin{eqnarray}
F(\alpha'\beta'\alpha\beta:\lambda_1\lambda_2\lambda_1'\lambda_2')&=&
\delta_{\alpha\lambda_1'}[{\cal A}(n^0_{\alpha'\lambda_1}n^0_{\beta'\lambda_2})n^0_{\lambda_2'\beta}
+n^0_{\lambda_2'\beta}C^0_{\alpha'\beta'\lambda_1\lambda_2}
+n^0_{\alpha'\lambda_1}C^0_{\beta'\lambda_2'\lambda_2\beta}
-n^0_{\beta'\lambda_1}C^0_{\alpha'\lambda_2'\lambda_2\beta}
\nonumber \\
&-&n^0_{\alpha'\lambda_2}C^0_{\beta'\lambda_2'\lambda_1\beta}
+n^0_{\beta'\lambda_2}C^0_{\alpha'\lambda_2'\lambda_1\beta}+C^0_{\alpha'\beta'\lambda_2'\lambda_1\lambda_2\beta}].
\label{F-func}
\end{eqnarray}
The three-body correlation matrix is also included in eq.(\ref{F-func}).
The matrix elements $A, B, C$, and $D$ are given in the following. 
For simplicity, terms containing $C^0_{\alpha\beta\gamma\alpha'\beta'\gamma'}$
are not shown. They appear in $B, C$, and $D$: Wherever there is a term containing 
$n^0_{\alpha\alpha'}C^0_{\beta\gamma\beta'\gamma'}$, there exists a corresponding term with $C^0_{\alpha\beta\gamma\alpha'\beta'\gamma'}$.
The expressions for $A, B, C$, and $D$ are
not unique and their symmetry properties are not necessarily apparent. 
Equations (\ref{grc1}), (\ref{grc2}), and (\ref{grc3}) allow us to take other expressions
and guarantee symmetry properties. 

\begin{eqnarray}
A(\alpha'\alpha:\lambda\lambda')&=&
(\epsilon_{\alpha'}-\epsilon_{\alpha})(\delta_{\lambda\alpha'}n^0_{\lambda'\alpha}-\delta_{\lambda'\alpha}n^0_{\alpha'\lambda})
\nonumber \\
&+&\sum_{\gamma\delta}[\langle\gamma\delta|v|\alpha\lambda\rangle
({\cal A}(n^0_{\lambda'\gamma}n^0_{\alpha'\delta})+C^0_{\lambda'\alpha'\gamma\delta})
+\langle\lambda'\alpha'|v|\delta\gamma\rangle
({\cal A}(n^0_{\gamma\lambda}n^0_{\delta\alpha})+C^0_{\delta\gamma\alpha\lambda})
\nonumber \\
&-&\langle\lambda'\delta|v|\alpha\gamma\rangle_A (n^0_{\gamma\lambda}n^0_{\alpha'\delta}
+C^0_{\alpha'\gamma\delta\lambda})
-\langle\gamma\alpha'|v|\delta\lambda\rangle_A(n^0_{\lambda'\gamma}n^0_{\delta\alpha}
+C^0_{\lambda'\delta\gamma\alpha})]
\nonumber \\
&-&\sum_{\gamma\delta\gamma'}(\langle\alpha'\gamma|v|\delta\gamma'\rangle
\delta_{\lambda'\alpha}C^0_{\delta\gamma'\lambda\gamma}
+\langle\gamma\delta|v|\alpha\gamma'\rangle\delta_{\lambda\alpha'}C^0_{\lambda'\gamma'\gamma\delta}),
\label{BA}
\\
B(\alpha'\beta'\alpha\beta:\lambda\lambda')&=&
\sum_{\gamma}\{\langle\lambda'\gamma|v|\alpha\beta\rangle_A[{\cal A}(n^0_{\alpha'\lambda}n^0_{\beta'\gamma})
+C^0_{\alpha'\beta'\lambda\gamma}]
+\langle\alpha'\beta'|v|\lambda\gamma\rangle_A[{\cal A}(n^0_{\lambda'\alpha}n^0_{\gamma\beta})
+C^0_{\lambda'\gamma\alpha\beta}]\}
\nonumber \\
&+&H(\alpha'\beta'\alpha\beta:\lambda\lambda') 
-H(\beta'\alpha'\alpha\beta:\lambda\lambda')
+H^*(\alpha\beta\alpha'\beta':\lambda'\lambda)
-H^*(\beta\alpha\alpha'\beta':\lambda'\lambda)
\nonumber \\
&+&I(\alpha'\beta'\alpha\beta:\lambda\lambda') - I(\alpha'\beta'\beta\alpha:\lambda\lambda') 
+I^*(\alpha\beta\alpha'\beta':\lambda'\lambda)-I^*(\alpha\beta\beta'\alpha':\lambda'\lambda),
\label{BB}\\
C(\alpha'\alpha:\lambda_1\lambda_2\lambda_1'\lambda_2')&=&
B(\lambda_1'\lambda_2'\lambda_1\lambda_2:\alpha\alpha'),
\label{BC}
\end{eqnarray}
where
\begin{eqnarray}
H(\alpha'\beta'\alpha\beta:\lambda\lambda')&=&-\delta_{\lambda\alpha'}\{
(\epsilon_{\alpha}+\epsilon_{\beta}-\epsilon_{\alpha'}-\epsilon_{\beta'})C^0_{\lambda'\beta'\alpha\beta}
\nonumber \\
&+&\sum_{\gamma\delta}(\langle\gamma\delta|v|\alpha\beta\rangle
({\cal A}(n^0_{\lambda'\gamma}n^0_{\beta'\delta})+C^0_{\lambda'\beta'\gamma\delta})
\nonumber \\
&-&\sum_{\gamma\delta\gamma'}[\langle\gamma\delta|v|\alpha\gamma'\rangle_A
(n^0_{\lambda'\gamma}n^0_{\beta'\delta}n^0_{\gamma'\beta}
+n^0_{\lambda'\gamma}C^0_{\beta'\gamma'\delta\beta}-n^0_{\beta'\gamma}C^0_{\lambda'\gamma'\delta\beta})
\nonumber \\
&+&\langle\gamma\delta|v|\alpha\gamma'\rangle 
n^0_{\gamma'\beta}C^0_{\lambda'\beta'\gamma\delta}
\nonumber \\
&-&\langle\gamma\delta|v|\beta\gamma'\rangle_A
(n^0_{\lambda'\gamma}n^0_{\beta'\delta}n^0_{\gamma'\alpha}
+n^0_{\lambda'\gamma}C^0_{\beta'\gamma'\delta\alpha}-n^0_{\beta'\gamma}C^0_{\lambda'\gamma'\delta\alpha})
\nonumber \\
&-&\langle\gamma\delta|v|\beta\gamma'\rangle 
n^0_{\gamma'\alpha}C^0_{\lambda'\beta'\gamma\delta}
\nonumber \\
&+&\langle\beta'\gamma|v|\delta\gamma'\rangle_A
(n^0_{\lambda'\gamma}n^0_{\delta\beta}n^0_{\gamma'\alpha}
+n^0_{\gamma'\alpha}C^0_{\delta\lambda'\beta\gamma}-n^0_{\gamma'\beta}C^0_{\delta\lambda'\alpha\gamma})
\nonumber \\
&+&\langle\beta'\gamma|v|\delta\gamma'\rangle n^0_{\lambda'\gamma}C^0_{\gamma'\delta\alpha\beta}]\},
\\
I(\alpha'\beta'\alpha\beta:\lambda\lambda')&=&
\sum_{\gamma\delta}\{
[\langle\gamma\delta|v|\alpha\lambda\rangle_A(
n^0_{\alpha'\gamma}n^0_{\beta'\delta}n^0_{\lambda'\beta}
+n^0_{\alpha'\gamma}C^0_{\beta'\lambda'\delta\beta}
-n^0_{\beta'\gamma}C^0_{\alpha'\lambda'\delta\beta})
\nonumber \\
&+&\langle\gamma\delta|v|\alpha\lambda\rangle
(n^0_{\lambda'\beta}C^0_{\alpha'\beta'\gamma\delta}
+n^0_{\lambda'\gamma}C^0_{\alpha'\beta'\delta\beta})]
\nonumber \\
&-&\langle\lambda'\gamma|v|\alpha\delta\rangle_A
[{\cal A}(n^0_{\alpha'\lambda}n^0_{\beta'\gamma})n^0_{\delta\beta}
+n^0_{\delta\lambda}C^0_{\alpha'\beta'\gamma\beta}
+n^0_{\delta\beta}C^0_{\alpha'\beta'\lambda\gamma}
\nonumber \\
&+&n^0_{\alpha'\lambda}C^0_{\beta'\delta\gamma\beta}
-n^0_{\beta'\lambda}C^0_{\alpha'\delta\gamma\beta}
-n^0_{\alpha'\gamma}C^0_{\beta'\delta\lambda\beta}
+n^0_{\beta'\gamma}C^0_{\alpha'\delta\lambda\beta}]\}.
\end{eqnarray}
The matrix $D$ is given as
\begin{eqnarray}
D(\alpha'\beta'\alpha\beta:\lambda_1\lambda_2\lambda_1'\lambda_2')&=&
(\epsilon_{\alpha'}+\epsilon_{\beta'}
-\epsilon_{\alpha}-\epsilon_{\beta})S_2(\alpha'\beta'\alpha\beta:\lambda_1\lambda_2\lambda_1'\lambda_2')
\nonumber \\
&+&
\langle\lambda_1'\lambda_2'|v|\alpha\beta\rangle_A({\cal A}
(n^0_{\alpha'\lambda_1}n^0_{\beta'\lambda_2})+C^0_{\alpha'\beta'\lambda_1\lambda_2})
+\langle\alpha'\beta'|v|\lambda_1\lambda_2\rangle_A({\cal A}(n^0_{\lambda_1'\alpha}n^0_{\lambda_2'\beta})+C^0_{\lambda_1'\lambda_2'\alpha\beta})
\nonumber \\
&+&J(\alpha'\beta'\alpha\beta:\lambda_1\lambda_2\lambda_1'\lambda_2')
+J^*(\alpha\beta\alpha'\beta':\lambda_1'\lambda_2'\lambda_1\lambda_2)
\nonumber \\
&+&K(\alpha'\beta'\alpha\beta:\lambda_1\lambda_2\lambda_1'\lambda_2')
+L(\alpha'\beta'\alpha\beta:\lambda_1\lambda_2\lambda_1'\lambda_2')
\nonumber \\
&+&{\rm all~other~exchange~terms~of~}K{\rm ~and~}L ,
\label{BD}
\end{eqnarray}
where
\begin{eqnarray}
J(\alpha'\beta'\alpha\beta:\lambda_1\lambda_2\lambda_1'\lambda_2')
&=&-{\cal A}(\delta_{\lambda_1\alpha'}\delta_{\lambda_2\beta'})\{
\sum_{\gamma\delta}[\langle\gamma\delta|v|\alpha\beta\rangle - 
\sum_{\gamma'}(\langle\gamma\delta|v|\alpha\gamma'\rangle n^0_{\gamma'\beta}
\nonumber \\
&-&\langle\gamma\delta|v|\beta\gamma'\rangle n^0_{\gamma'\alpha}))]
({\cal A}(n^0_{\lambda_1'\gamma}n^0_{\lambda_2'\delta})+C^0_{\lambda_1'\lambda_2'\gamma\delta})
\nonumber \\
&+&\sum_{\gamma\delta\gamma'}[\langle\gamma\delta|v|\alpha\gamma'\rangle 
(n^0_{\lambda_1'\beta}C^0_{\gamma'\lambda_2'\gamma\delta}
-n^0_{\lambda_2'\beta}C^0_{\gamma'\lambda_1'\gamma\delta}
-n^0_{\lambda_1'\gamma}C^0_{\gamma'\lambda_2'\beta\delta}
+n^0_{\lambda_2'\gamma}C^0_{\gamma'\lambda_1'\beta\delta})
\nonumber \\
&-&\langle\gamma\delta|v|\beta\gamma'\rangle 
(n^0_{\lambda_1'\alpha}C^0_{\gamma'\lambda_2'\gamma\delta}
-n^0_{\lambda_2'\alpha}C^0_{\gamma'\lambda_1'\gamma\delta}
-n^0_{\lambda_1'\gamma}C^0_{\gamma'\lambda_2'\alpha\delta}
+n^0_{\lambda_2'\gamma}C^0_{\gamma'\lambda_1'\alpha\delta})]\},
\\
K(\alpha'\beta'\alpha\beta:\lambda_1\lambda_2\lambda_1'\lambda_2')&=&
\delta_{\lambda_1\alpha'}\{
\sum_{\gamma\delta}[\langle\gamma\delta|v|\alpha\beta\rangle_A(
n^0_{\lambda_1'\gamma}n^0_{\lambda_2'\delta}n^0_{\beta'\lambda_2}
+n^0_{\lambda_1'\gamma}C^0_{\lambda_2'\beta'\delta\lambda_2})
\nonumber \\
&+&\langle\gamma\delta|v|\alpha\lambda_2\rangle_A(
n^0_{\lambda_1'\gamma}n^0_{\lambda_2'\beta}n^0_{\beta'\delta}
+n^0_{\lambda_1'\gamma}C^0_{\lambda_2'\beta'\beta\delta})
\nonumber \\
&+&\langle\gamma\beta'|v|\lambda_2\delta\rangle_A(n^0_{\delta\beta}
n^0_{\lambda_1'\alpha}n^0_{\lambda_2'\gamma}+n^0_{\delta\beta}C^0_{\lambda_1'\lambda_2'\alpha\gamma})
\nonumber \\
&+&{\rm all~other~exchange~terms }]
\nonumber \\
&-&\sum_{\gamma\delta\gamma'}[
\langle\gamma\delta|v|\alpha\gamma'\rangle_A(
n^0_{\lambda_1'\gamma}n^0_{\beta'\delta}n^0_{\lambda_2'\beta}n^0_{\gamma'\lambda_2}
+n^0_{\lambda_1'\gamma}n^0_{\beta'\delta}C^0_{\lambda_2'\gamma'\beta\lambda_2}
+C^0_{\lambda_1'\beta'\gamma\delta}C^0_{\lambda_2'\gamma'\beta\lambda_2})
\nonumber \\
&-&\langle\gamma\beta'|v|\gamma'\delta\rangle_A(
n^0_{\lambda_1'\alpha}n^0_{\lambda_2'\gamma}n^0_{\gamma'\beta}n^0_{\delta\lambda_2}
+n^0_{\gamma'\beta}n^0_{\lambda_1'\alpha}C^0_{\lambda_2'\delta\gamma\lambda_2}
+C^0_{\lambda_2'\gamma'\alpha\beta}C^0_{\lambda_1'\delta\gamma\lambda_2})
\nonumber \\
&+&{\rm all~other~exchange~terms }]\},
\\
L(\alpha'\beta'\alpha\beta:\lambda_1\lambda_2\lambda_1'\lambda_2')&=&
-\sum_{\gamma}[\langle\alpha'\gamma|v|\lambda_1\lambda_2\rangle (n^0_{\beta'\gamma}
n^0_{\lambda_1'\alpha}n^0_{\lambda_2'\beta}+n^0_{\beta'\gamma}C^0_{\lambda_1'\lambda_2'\alpha\beta})
\nonumber \\
&+&{\rm all~other~exchange~terms }]
\nonumber \\
&+&\sum_{\gamma\delta}[\langle\alpha'\gamma|v|\lambda_1\delta\rangle_A
(n^0_{\delta\beta}n^0_{\lambda_1'\alpha}n^0_{\lambda_2'\gamma}n^0_{\beta'\lambda_2}
+n^0_{\beta'\lambda_2}n^0_{\delta\beta}C^0_{\lambda_1'\lambda_2'\alpha\gamma}
+C^0_{\beta'\delta\lambda_2\beta}C^0_{\lambda_1'\lambda_2'\alpha\gamma})
\nonumber \\
&+&{\rm all~other~exchange~terms }].
\end{eqnarray}
Finally we discuss a relation between eq.(\ref{erpa3}) and a set of the STDDM equations (eqs.(\ref{stddm1}) and (\ref{stddm2})).
When the ground state is approximated by the HF one, eqs.(\ref{BS1})-(\ref{BS2}) become
\begin{eqnarray}
S_1(\alpha'\alpha:\lambda\lambda')
&=&(f_{\alpha}-f_{\alpha'})\delta_{\alpha\lambda'}\delta_{\alpha'\lambda},\\
S_2(\alpha'\beta'\alpha\beta:\lambda_1\lambda_2\lambda_1'\lambda_2')
&=& {\cal A}(\delta_{\alpha\lambda_1'}\delta_{\beta\lambda_2'}){\cal A}(\delta_{\alpha'\lambda_1}\delta_{\beta'\lambda_2})F^0_{\alpha'\beta'\alpha\beta},
\\
T_1(\alpha'\alpha:\lambda_1\lambda_2\lambda_1'\lambda_2')
&=&
T_2(\alpha'\beta'\alpha\beta:\lambda\lambda')
=0 ,
\end{eqnarray}
where
\begin{eqnarray}
F^0_{\alpha'\beta'\alpha\beta}=
f_{\alpha}f_{\beta}\bar{f}_{\alpha'}\bar{f}_{\beta'}
-\bar{f}_{\alpha}\bar{f}_{\beta}f_{\alpha'}f_{\beta'}.
\end{eqnarray}
Equations (\ref{BA})-(\ref{BC}) and (\ref{BD}) become the following:
\begin{eqnarray}
A(\alpha'\alpha:\lambda\lambda')&=&
[(\epsilon_{\alpha'}-\epsilon_{\alpha})\delta_{\alpha\lambda'}\delta_{\alpha'\lambda}
+\langle\lambda'\alpha'|v|\alpha\lambda\rangle_A(f_{\alpha'}-f_{\alpha})](f_{\lambda'}-f_{\lambda}),
\label{BA1}
\\
B(\alpha'\beta'\alpha\beta:\lambda\lambda')&=&-[
(\bar{f_{\alpha}}\bar{f_{\beta}}f_{\beta'}
+f_{\alpha}f_{\beta}\bar{f}_{\beta'})\langle\lambda'\beta'|v|\alpha\beta\rangle_A\delta_{\alpha'\lambda}
-(\bar{f_{\alpha}}\bar{f_{\beta}}f_{\alpha'}+f_{\alpha}f_{\beta}
\bar{f}_{\alpha'})\langle\lambda'\alpha'|v|\alpha\beta\rangle_A\delta_{\beta'\lambda}
\nonumber \\
&-&(\bar{f_{\beta}}f_{\alpha'}f_{\beta'}
+f_{\beta}\bar{f}_{\alpha'}\bar{f}_{\beta'})\langle\alpha'\beta'|v|\lambda\beta\rangle_A\delta_{\alpha\lambda'}
\nonumber \\
&+&(\bar{f_{\alpha}}f_{\alpha'}f_{\beta'}+f_{\alpha}\bar{f}_{\alpha'}
\bar{f}_{\beta'})\langle\alpha'\beta'|v|\lambda\alpha\rangle_A\delta_{\beta\lambda'}](f_{\lambda'}-f_{\lambda}),
\label{BB1}
\\
C(\alpha'\alpha:\lambda_1\lambda_2\lambda_1'\lambda_2')&=&
B(\lambda_1'\lambda_2'\lambda_1\lambda_2:\alpha\alpha'),
\\
D(\alpha'\beta'\alpha\beta:\lambda_1\lambda_2\lambda_1'\lambda_2')&=&
F^0_{\lambda_1\lambda_2\lambda_1'\lambda_2'}\{
(\epsilon_{\alpha'}+\epsilon_{\beta'}
-\epsilon_{\alpha}-\epsilon_{\beta})
{\cal A}(\delta_{\alpha\lambda_1'}\delta_{\beta\lambda_2'}){\cal A}(\delta_{\alpha'\lambda_1}\delta_{\beta'\lambda_2})
\nonumber \\
&+&(1-f_{\alpha'}-f_{\beta'})
\langle\alpha'\beta'|v|\lambda_1\lambda_2\rangle_A{\cal A}(\delta_{\alpha\lambda_1'}\delta_{\beta\lambda_2'})
\nonumber \\
&-&(1-f_{\alpha}-f_{\beta})
\langle\lambda_1'\lambda_2'|v|\alpha\beta\rangle_A{\cal A}(\delta_{\alpha'\lambda_1}\delta_{\beta'\lambda_2})
\nonumber \\
&+&(f_{\alpha}-f_{\alpha'})[
\langle\alpha'\lambda_1'|v|\alpha\lambda_1\rangle_A\delta_{\beta'\lambda_2}\delta_{\beta\lambda_2'}
+\langle\alpha'\lambda_2'|v|\alpha\lambda_2\rangle_A\delta_{\beta'\lambda_1}\delta_{\beta\lambda_1'}
\nonumber \\
&-&\langle\alpha'\lambda_1'|v|\alpha\lambda_2\rangle_A\delta_{\beta'\lambda_1}\delta_{\beta\lambda_2'}
-\langle\alpha'\lambda_2'|v|\alpha\lambda_1\rangle_A\delta_{\beta'\lambda_2}\delta_{\beta\lambda_1'}]
\nonumber \\
&+&(f_{\beta}-f_{\beta'})[
\langle\beta'\lambda_1'|v|\beta\lambda_1\rangle_A\delta_{\alpha'\lambda_2}\delta_{\alpha\lambda_2'}
+\langle\beta'\lambda_2'|v|\beta\lambda_2\rangle_A\delta_{\alpha'\lambda_1}\delta_{\alpha\lambda_1'}
\nonumber \\
&-&\langle\beta'\lambda_1'|v|\beta\lambda_2\rangle_A\delta_{\alpha'\lambda_1}\delta_{\alpha\lambda_2'}
-\langle\beta'\lambda_2'|v|\beta\lambda_1\rangle_A\delta_{\alpha'\lambda_2}\delta_{\alpha\lambda_1'}]
\nonumber \\
&-&(f_{\alpha}-f_{\beta'})[
\langle\beta'\lambda_1'|v|\alpha\lambda_1\rangle_A\delta_{\alpha'\lambda_2}\delta_{\beta\lambda_2'}
+\langle\beta'\lambda_2'|v|\alpha\lambda_2\rangle_A\delta_{\alpha'\lambda_1}\delta_{\beta\lambda_1'}
\nonumber \\
&-&\langle\beta'\lambda_1'|v|\alpha\lambda_2\rangle_A\delta_{\alpha'\lambda_1}\delta_{\beta\lambda_2'}
-\langle\beta'\lambda_2'|v|\alpha\lambda_1\rangle_A\delta_{\alpha'\lambda_2}\delta_{\beta\lambda_1'}]
\nonumber \\
&-&(f_{\alpha}-f_{\beta'})[
\langle\beta'\lambda_1'|v|\alpha\lambda_1\rangle_A\delta_{\alpha'\lambda_2}\delta_{\beta\lambda_2'}
+\langle\beta'\lambda_2'|v|\alpha\lambda_2\rangle_A\delta_{\alpha'\lambda_1}\delta_{\beta\lambda_1'}
\nonumber \\
&-&\langle\beta'\lambda_1'|v|\alpha\lambda_2\rangle_A\delta_{\alpha'\lambda_1}\delta_{\beta\lambda_2'}
-\langle\beta'\lambda_2'|v|\alpha\lambda_1\rangle_A\delta_{\alpha'\lambda_2}\delta_{\beta\lambda_1'}]\}.
\end{eqnarray}
If $S_1x$ and $S_2X$ which appear in the equation for $X$, that is, $Bx+DX=\omega S_2X$, are replaced by $x$ and $X$, respectively, the equation 
for $X$ is equivalent to eq.(\ref{stddm2}). However, the replacement $S_1x\rightarrow x$ and
$S_2X\rightarrow X$ in 
$Ax+CX=\omega S_1x$ cannot give eq.(\ref{stddm1}) because of the symmetric coupling between $x$ and $X$. 
Since the expression for
$C$ (eq.(\ref{BC}) is not unique as mentioned above, 
we can always take an expression for $C$ which leads
to the same coupling matrix as in eq.(\ref{stddm1}) in the HF limit. Such an expression for $C$ is the following:
\begin{eqnarray}
C(\alpha'\alpha:\lambda_1\lambda_2\lambda_1'\lambda_2')&=&F^0_{\lambda_1\lambda_2\lambda_1'\lambda_2'}(
\langle\alpha'\lambda_2'|v|\lambda_1\lambda_2\rangle_A\delta_{\lambda_1'\alpha}
-\langle\alpha'\lambda_1'|v|\lambda_1\lambda_2\rangle_A\delta_{\lambda_2'\alpha}
\nonumber \\
&-&\langle\lambda_1'\lambda_2'|v|\alpha\lambda_2\rangle_A\delta_{\lambda_1\alpha'}
+\langle\lambda_1'\lambda_2'|v|\alpha\lambda_1\rangle_A\delta_{\lambda_2\alpha'}).
\end{eqnarray}

\end{document}